\documentclass[preprintnumbers]{revtex4}
\usepackage{amsfonts}
\usepackage{amsmath}
\usepackage{hyperref}
\usepackage{amssymb}
\usepackage[english]{babel}
\usepackage{graphicx}
\usepackage{epsfig}
\usepackage{bm}
\usepackage{longtable}
\usepackage{color}
\usepackage{verbatim}
\usepackage{longtable}
\usepackage[utf8]{inputenc}

\newcommand{\mathsym}[1]{{}}

\input{tcilatex}

\begin{document}

\title{A 3-3-1 model with right-handed neutrinos based on the $\Delta \left( 27\right) $ family symmetry.}
\author{A. E. C\'{a}rcamo Hern\'{a}ndez${}^{a}$}
\email{antonio.carcamo@usm.cl}
\author{H. N. Long$^{b}$}
\email{hnlong@iop.vast.ac.vn}
\author{V. V. Vien$^{c}$}
\email{wvienk16@gmail.com}
\affiliation{$^{{a}}$Universidad T\'{e}cnica Federico Santa Mar\'{\i}a\\
and Centro Cient\'{\i}fico-Tecnol\'{o}gico de Valpara\'{\i}so\\
Casilla 110-V, Valpara\'{\i}so, Chile,}
\affiliation{$^{{b}}$Institute of Physics, Vietnam Academy of Science and Technology, \\
10 Dao Tan, Ba Dinh, Hanoi, Vietnam}
\affiliation{${}^{c}$ Institute of Research and Development, Duy Tan University, 182
Nguyen Van Linh, Da Nang City, Vietnam \\
and Department of Physics, Tay Nguyen University, 567 Le Duan, Buon Ma
Thuot, DakLak, Vietnam}
\date{\today }

\begin{abstract}
We present the first multiscalar singlet extension of the 3-3-1 model with right-handed neutrinos, based on the $\Delta \left( 27\right) $ family symmetry, supplemented by the $Z_{4}\otimes Z_{8}\otimes Z_{14}$ flavor group, consistent with
 current low energy fermion flavor data. In the model under consideration, the light active neutrino masses are
generated from a double seesaw mechanism and the observed pattern of charged
fermion masses and quark mixing angles is caused by the breaking of the $\Delta \left( 27\right) \otimes Z_{4}\otimes Z_{8}\otimes Z_{14}$ discrete group at very high energy. Our model has only 14 effective free parameters, which are fitted to reproduce the experimental values of the 18 physical observables in the quark and lepton sectors. The obtained physical observables for the quark sector agree with their experimental values, whereas those ones for the lepton sector also do, only for the inverted neutrino mass hierarchy. The normal neutrino mass hierarchy scenario of the model is disfavored by the neutrino oscillation experimental data. We find an effective Majorana neutrino mass parameter of neutrinoless double beta decay of $m_{\beta \beta }=$ 22 meV, a leptonic Dirac CP violating phase of $34^{\circ }$ and a Jarlskog invariant of about $10^{-2}$ for the inverted neutrino mass spectrum.
\end{abstract}

\maketitle
\section{Introduction}
The observation of the $125$ GeV Higgs boson at the LHC \cite{Aad:2012tfa,Chatrchyan:2012xdj}, confirmed the great success of the
Standard Model (SM) as the right theory of electroweak interactions. Despite
the couplings of this scalar state with the SM particles are very consistent
with the properties expected of the SM Higgs boson, the possibility that new
scalar states may exist and play a role in the Electroweak Symmetry Breaking
(EWSB) mechanism, is still open. The current priority of the LHC experiments
will be to make very precise measurements of the Higgs boson selfcouplings
as well as of its couplings to the SM particles with the aim to shed light
on the underlying theory behind Electroweak Symmetry Breaking (EWSB).
Furthermore, despite its great experimental success, there are several
aspects not explained in the context of the SM, such as, for example, the
smallness of neutrino masses, the observed pattern of fermion masses and
mixing angles and the existence of three generations of fermions. The SM
does not explain why in the quark sector the mixing angles are small,
whereas in the lepton sector two of the mixing angles are large and one is
small. The Daya Bay \cite{An:2012eh}, T2K \cite{Abe:2011sj}, MINOS \cite%
{Adamson:2011qu}, Double CHOOZ \cite{Abe:2011fz} and RENO \cite{Ahn:2012nd}
neutrino oscillation experiments, clearly indicate that at least two of the
light active neutrinos have non-vanishing masses. These experiments have
provided important constraints on the neutrino mass squared splittings and
leptonic mixing parameters \cite{Forero:2014bxa}. Furthermore, the SM does
not provide an explanation for the charged fermion mass hierarchy, which is
extended over a range of about 11 orders of magnitude, from the neutrino
mass scale up to the top quark mass.

\quad The unexplained SM fermion mass and mixing pattern motivates us to
consider models with extended symmetry and larger scalar and/or fermion
content, addressed to explain the fermion mass and mixing pattern. There are
two approaches to describe the observed fermion mass and mixing pattern:
assuming Yukawa textures \cite%
{Fritzsch:1977za,Fukuyama:1997ky,Du:1992iy,Barbieri:1994kw,Peccei:1995fg,Fritzsch:1999ee,Roberts:2001zy,Nishiura:2002ei,deMedeirosVarzielas:2005ax,Carcamo:2006dp,Kajiyama:2007gx,CarcamoHernandez:2010im,Branco:2010tx,Leser:2011fz,Gupta:2012dma,CarcamoHernandez:2012xy,Hernandez:2013mcf,Pas:2014bra,Hernandez:2014hka,Hernandez:2014zsa,Nishiura:2014psa,Frank:2014aca,Sinha:2015ooa,Nishiura:2015qia,Gautam:2015kya,Pas:2015hca,Hernandez:2015hrt}
and implementing discrete flavor groups in extensions of the SM (see Refs. 
\cite{Ishimori:2010au,Altarelli:2010gt,King:2013eh, King:2014nza} for recent
reviews on flavor symmetries). Recently, extensions of the SM with $A_{4}$%
\cite%
{Ma:2001dn,He:2006dk,Chen:2009um,Dong:2010gk,Ahn:2012tv,Memenga:2013vc,Felipe:2013vwa,Varzielas:2012ai,Ishimori:2012fg,King:2013hj,Hernandez:2013dta,Babu:2002dz,Altarelli:2005yx,Morisi:2013eca,Altarelli:2005yp,Kadosh:2010rm,Kadosh:2013nra,delAguila:2010vg,Campos:2014lla,Vien:2014pta,Hernandez:2015tna,Nishi:2016jqg}%
, $S_{3}$ \cite%
{Chen:2004rr,Dong:2011vb,Bhattacharyya:2010hp,Dias:2012bh,Meloni:2012ci,Canales:2013cga,Ma:2013zca,Kajiyama:2013sza,Hernandez:2013hea,Ma:2014qra,Hernandez:2014vta,Hernandez:2014lpa,Hernandez:2015dga,Hernandez:2015zeh,Hernandez:2016rbi}%
, $S_{4}$ \cite%
{Mohapatra:2012tb,BhupalDev:2012nm,Varzielas:2012pa,Ding:2013hpa,Ishimori:2010fs,Ding:2013eca,Hagedorn:2011un,Campos:2014zaa,Dong:2010zu,VanVien:2015xha}%
, $D_4$ \cite%
{Frampton:1994rk,Grimus:2003kq,Grimus:2004rj,Frigerio:2004jg,Babu:2004tn,Adulpravitchai:2008yp,Ishimori:2008gp,Hagedorn:2010mq,Meloni:2011cc,Vien:2013zra}%
, $T_7$ \cite%
{Luhn:2007sy,Hagedorn:2008bc,Cao:2010mp,Luhn:2012bc,Kajiyama:2013lja,Bonilla:2014xla,Vien:2014gza,Vien:2015koa,Hernandez:2015cra,Arbelaez:2015toa}%
, $T_{13}$ \cite{Ding:2011qt,Hartmann:2011dn,Hartmann:2011pq,Kajiyama:2010sb}%
, $T^{\prime }$ \cite%
{Aranda:2000tm,Aranda:2007dp,Chen:2007afa,Frampton:2008bz,Eby:2011ph,Frampton:2013lva}
and $\Delta(27)$ \cite%
{Ma:2007wu,Varzielas:2012nn,Bhattacharyya:2012pi,Ma:2013xqa,Nishi:2013jqa,Varzielas:2013sla,Aranda:2013gga,Ma:2014eka,Abbas:2014ewa,Abbas:2015zna,Varzielas:2015aua,Bjorkeroth:2015uou,Chen:2015jta,Vien:2016tmh}
family symmetries have been considered to address the flavor puzzle of the
SM.

\quad On the other hand, the existence of three fermion families, which is
not explained in the context of the SM, can be understood in the framework
of models with $SU(3)_{C}\otimes SU(3)_{L}\otimes U(1)_{X}$ gauge symmetry,
called 3-3-1 models for short, where $U(1)_{X}$ is a nonuniversal family
symmetry that distinguishes the third fermion family from the first and
second ones \cite%
{Georgi:1978bv,Valle:1983dk,Pisano:1991ee,Montero:1992jk,Foot:1992rh,Frampton:1992wt,Ng:1992st,Duong:1993zn,Hoang:1996gi,Hoang:1995vq,Foot:1994ym,Diaz:2003dk,Diaz:2004fs,Dias:2004dc,Dias:2005yh,Dias:2005jm,Ochoa:2005ch,CarcamoHernandez:2005ka,Dias:2010vt,Dias:2012xp,Alvarado:2012xi,Catano:2012kw,Hernandez:2013mcf,Hernandez:2014lpa,Vien:2014pta,Hernandez:2014vta,Boucenna:2014ela,Boucenna:2014dia,Vien:2014gza,Phong:2014ofa,Boucenna:2015zwa,Hernandez:2015cra,DeConto:2015eia,Correia:2015tra,Hernandez:2015tna,Okada:2015bxa,Binh:2015cba,Hue:2015fbb}%
. These models have several interesting features. First, the existence of
three generations of fermions is a consequence of the chiral anomaly
cancellation and the asymptotic freedom in QCD. Second, the large mass
hierarchy between the heaviest quark family and the two lighter ones can be
understood from the fact that the former has a different $U(1)_{X}$ charge
from the latter. Third, these models include a natural Peccei-Quinn
symmetry, which addresses the strong-CP problem \cite%
{Pal:1994ba,Dias:2002gg,Dias:2003zt,Dias:2003iq}. Finally, versions with
heavy sterile neutrinos include cold dark matter candidates as weakly
interacting massive particles (WIMPs) \cite{Mizukoshi:2010ky}. Besides that,
the 3-3-1 models can explain the $750$ GeV diphoton excess recently reported
by ATLAS and CMS \cite{Boucenna:2015pav,Hernandez:2015ywg,Dong:2015dxw,Cao:2015scs} as well as the $2$ TeV diboson excess found by ATLAS \cite{Cao:2015lia}.

\quad In the 3-3-1 models, the electroweak gauge symmetry is broken in two
steps as follows. First the $SU(3)_{L}\otimes U(1)_{X}$ symmetry is broken
down to the SM electroweak group $SU(2)_{L}\otimes U(1)_{Y}$ by one heavy $%
SU(3)_L$ triplet field acquiring a Vacuum Expectation Value (VEV) at high
energy scale $v _{\chi }$, thus giving masses to non SM fermions and gauge
bosons. Second, the usual EWSB mechanism is
triggered by the remaining lighter triplets, with VEVs at the electroweak
scale $\upsilon _{\rho }$ and $v_{\eta }$, thus providing SM fermions and gauge bosons with masses \cite{Hernandez:2013mcf}.

\quad In Ref. \cite{Vien:2016tmh} we have proposed a 3-3-1 model with $%
\Delta \left( 27\right) $ flavor symmetry supplemented by the $\mathrm{U}%
(1)_{\mathcal{L}}$ new lepton global symmetry that enforces to have
different scalar fields in the Yukawa interactions for charged lepton,
neutrino and quark sectors, thus allowing us to treat these sectors
independently. The scalar sector of that model includes 10 $SU(3)_L$ scalar
triplets and three $SU(3)_L$ scalar antisextets. The $\mathrm{SU}%
(3)_{C}\otimes \mathrm{SU}(3)_{L}\otimes \mathrm{U}(1)_{X}\otimes\mathrm{U}%
(1)_{\mathcal{L}}\otimes\Delta\left( 27\right)$ assignments of the fermion
sector of our previous model, require that these 10 $SU(3)_L$ scalar
triplets be distributed as follows, 4 for the quark sector, 3 for the
charged lepton sector and 3 for the neutrino sector. Furthermore the 3 $%
SU(3)_L$ scalar antisextets are needed to implement a type I seesaw
mechanism. In that model, light active neutrino masses are generated from
type-I and type-II seesaw mechanisms, mediated by three heavy right handed
Majorana neutrinos and three $SU(3)_{L}$ scalar antisextets, respectively.
Since the Yukawa terms in that model are renormalizable, to explain the
charged fermion mass pattern, one needs to impose a strong hierarchy among
the charged fermion Yukawa couplings of the model.

\quad It is interesting to find an alternative and better explanation for
the SM fermion mass and mixing hierarchy, by formulating a 3-3-1 model with
less scalar content than our previous model of Ref. \cite{Vien:2016tmh}. To
this end, we propose an alternative and improved version of the 3-3-1 model
based on the $\mathrm{SU}(3)_{C}\otimes \mathrm{SU}(3)_{L}\otimes \mathrm{U}%
(1)_{X}\otimes \Delta \left( 27\right) \otimes Z_{4}\otimes Z_{8}\otimes
Z_{14}$ symmetry that successfully describes the observed fermion mass and
mixing pattern and is consistent with the current low energy fermion flavor
data. The particular role of each discrete group factor is explained in
detail in Sec. \ref{model}. The scalar sector of our model includes 3 $SU(3)_L$ scalar triplets and 22 $SU(3)_L$ scalar singlets, assigned into
triplet and singlet irreducible representations of the $\Delta(27)$ discrete
group. This scalar sector of our current $\Delta(27)$ flavor 3-3-1 model is
more minimal than that one of our previous model of Ref. \cite{Vien:2016tmh}
and does not include $SU(3)_L$ scalar antisextets. Furthermore, our current
model does not include the $\mathrm{U}(1)_{\mathcal{L}}$ new lepton global
symmetry presented in our previous $\Delta(27)$ flavor 3-3-1 model. Unlike
our previous $\Delta(27)$ flavor 3-3-1 model of Ref. \cite{Vien:2016tmh}, in
our current 3-3-1 model, the charged fermion mass and quark mixing pattern
can successfully be accounted for, by having all Yukawa couplings of order
unity and arises from the breaking of the $\Delta \left( 27\right) \otimes
Z_{4}\otimes Z_{8}\otimes Z_{14}$ discrete group at very high energy,
triggered by $SU(3)_L$ scalar singlets acquiring vacuum expectation values
much larger than the TeV scale.

\quad In the following we summarize the most important differences of our current $\Delta(27)$ flavor 3-3-1 model with our previous 3-3-1 model also based on the $\Delta(27)$ family symmetry. First of all, the scalar sector of our current 3-3-1 model has 3 $SU(3)_L$ scalar triplets plus 22 very heavy $SU(3)_L$ scalar singlets. On the other hand, our previous $\Delta(27)$ flavor 3-3-1 model has a scalar sector composed of 10 $SU(3)_L$ scalar triplets and three $SU(3)_{L}$ scalar antisextets. Second, the charged fermion mass and quark mixing pattern can successfully be accounted for in our current 3-3-1 model with $\Delta(27)$ family symmetry by having the Yukawa couplings of order unity, whereas in our previous $\Delta(27)$ flavor 3-3-1 model, a strong hierarchy of the Yukawa couplings is needed to accommodate the current pattern of charged fermion masses and the CKM quark mixing matrix is predicted to be equal to the identity matrix. Third, in our current 3-3-1 model with $\Delta(27)$ family symmetry the light active neutrino masses arise from a double seesaw mechanism whereas in our previous $\Delta(27)$ flavor 3-3-1 model, type I and type II seesaw mechanisms generate the masses for the light active neutrinos. Finally, our current 3-3-1 model with $\Delta(27)$ family symmetry, does not include the $\mathrm{U}(1)_{\mathcal{L}}$ new lepton global symmetry presented in our previous $\Delta(27)$ flavor 3-3-1 model, but has instead a $Z_{4}\otimes Z_{8}\otimes Z_{14}$ discrete symmetry, whose breaking at very high energy gives rise to the observed pattern of charged fermion masses and quark mixing angles.

\quad It is noteworthy that our previous $\Delta(27)$ flavor 3-3-1 model model corresponds to an extension of the original 3-3-1 model with right handed Majorana neutrinos (which includes 3 $SU(3)_L$ scalar triplets in its scalar spectrum), where 7 extra $SU(3)_L$ scalar triplets and 3 $SU(3)_{L}$ scalar antisextets are added to build the charged fermion and neutrino Yukawa terms needed to give masses to SM charged fermions and light active neutrinos. On the other hand, in our current $\Delta(27)$ flavor 3-3-1 model, preserves the content of particles of the 3-3-1 model with right handed Majorana neutrinos, but we add additional very heavy $SU(3)_L$ singlet scalar fields with quantum numbers that allow to build Yukawa terms invariant under the local and discrete groups. Consequently our current model corresponds to the first multiscalar singlet extension of the original 3-3-1 model with right-handed neutrinos, based on the $\Delta(27)$ family symmetry.  As these singlet scalars fields are assumed to be much heavier than the 3 $SU(3)_L$ scalar triplets, our model at low energies reduces to the 3-3-1 model with right-handed neutrinos.

\quad In this paper we propose the first implementation of the $\Delta(27)$
flavor symmetry in a multiscalar singlet extension of the original 3-3-1 model with
right-handed neutrinos. In our model, light active neutrino masses arise
from a double seesaw mechanism mediated by three heavy right-handed Majorana
neutrinos. This paper is organized as follows. In Sect. \ref{model} we
outline the proposed model. In Sect. \ref{leptonmassesandmixing} we discuss
the implications of our model in masses and mixings in the lepton sector. In
Sect. \ref{quarkmassesandmixing} we present a discussion of quark masses and
mixings, followed by a numerical analysis. Finally we conclude in Sect. \ref{conclusions}. Appendix \ref{A} provides a description of the $\Delta \left(
27\right) $ discrete group. Appendix \ref{ApB} includes a discussion of the scalar potential for two $\Delta(27)$ scalar triplets and its minimization equations.

\section{The model}

\label{model}The first 3-3-1 model with right handed Majorana neutrinos in
the $SU(3)_{L}$ lepton triplet was considered in \cite{Montero:1992jk}.
However that model cannot describe the observed pattern of fermion masses
and mixings, due to the unexplained hierarchy among the large number of
Yukawa couplings in the model. Below we consider a multiscalar singlet
extension of the $SU(3)_{C}\otimes SU\left( 3\right) _{L}\otimes U\left(
1\right) _{X}$ (3-3-1) model with right-handed neutrinos, which successfully
describes the SM fermion mass and mixing pattern. In our model the full
symmetry $\mathcal{G}$ experiences the following three-step spontaneous
breaking: 
\begin{widetext}
\begin{eqnarray}
&&\mathcal{G}=SU(3)_{C}\otimes SU\left( 3\right) _{L}\otimes U\left(
1\right) _{X}\otimes \Delta \left( 27\right) \otimes Z_{4}\otimes
Z_{8}\otimes Z_{14}{\xrightarrow{\Lambda _{int}}}SU(3)_{C}\otimes SU\left(
3\right) _{L}\otimes U\left( 1\right) _{X}{\xrightarrow{v_{\chi }}}
\label{Group} \\
&&\hspace{15mm}SU(3)_{C}\otimes SU\left( 2\right) _{L}\otimes U\left(
1\right) _{Y}{\xrightarrow{v_{\eta },v_{\rho }}}SU(3)_{C}\otimes U\left(
1\right) _{Q},\notag
\end{eqnarray}
\end{widetext}
and the symmetry breaking scales obey the relation $\Lambda _{int}\gg
v_{\chi }\gg v_{\eta },v_{\rho }.$

We define the electric charge of our 3-3-1 model as a combination of the $%
SU(3)$ generators and the identity, as follows: 
\begin{equation}
Q=T_{3}-\frac{1}{\sqrt{3}}T_{8}+XI,
\end{equation}%
with $I=diag(1,1,1)$, $T_{3}=\frac{1}{2}diag(1,-1,0)$ and $T_{8}=(\frac{1}{2%
\sqrt{3}})diag(1,1,-2)$ for triplet.

From the requirement of anomaly cancellation, it follows that the fermions
of our model are assigned in the following $(SU(3)_{C},SU(3)_{L},U(1)_{X})$
left- and right-handed representations: 
\begin{align}
Q_{L}^{1,2}& =%
\begin{pmatrix}
D^{1,2} \\ 
-U^{1,2} \\ 
J^{1,2} \\ 
\end{pmatrix}%
_{L}:(3,3^{\ast },0),\left\{ 
\begin{array}{c}
D_{R}^{1,2}:(3,1,-1/3), \\ 
U_{R}^{1,2}:(3,1,2/3), \\ 
J_{R}^{1,2}:(3,1,-1/3), \\ 
\end{array}%
\right.  \notag \\
Q_{L}^{3}& =%
\begin{pmatrix}
U^{3} \\ 
D^{3} \\ 
T \\ 
\end{pmatrix}%
_{L}:(3,3,1/3),\left\{ 
\begin{array}{c}
U_{R}^{3}:(3,1,2/3), \\ 
D_{R}^{3}:(3,1,-1/3), \\ 
T_{R}:(3,1,2/3), \\ 
\end{array}%
\right.  \label{fermion_spectrum} \\
L_{L}^{1,2,3}& =%
\begin{pmatrix}
\nu ^{1,2,3} \\ 
e^{1,2,3} \\ 
(\nu ^{1,2,3})^{c} \\ 
\end{pmatrix}%
_{L}:(1,3,-1/3),\left\{ 
\begin{array}{c}
e_{R}^{1,2,3}:(1,1,-1), \\ 
N_{R}^{1,2,3}:(1,1,0), \\ 
\end{array}%
\right.\notag  
\end{align}%
where $U_{L}^{i}$ and $D_{L}^{i}$ for $i=1,2,3$ are three up- and down-type
quark components in the flavor basis, while $\nu _{L}^{i}$ and $e_{L}^{i}$
are the neutral and charged lepton families. The right-handed fermions
transform as singlets under $SU(3)_{L}$ with $U(1)_{X}$ quantum numbers
equal to their electric charges. Furthermore, the model has the following
heavy fermions: a single flavor quark $T$ with electric charge $2/3$, two
flavor quarks $J^{1,2}$ with charge $-1/3$, three neutral Majorana leptons $%
(\nu ^{1,2,3})_{L}^{c}$ and three right-handed Majorana leptons $%
N_{R}^{1,2,3}$.

Regarding the scalar sector of the 3-3-1 model with right handed Majorana
neutrinos, we assign the scalar fields in the following $%
[SU(3)_{L},U(1)_{X}] $ representations: 
\begin{align}
\chi & =%
\begin{pmatrix}
\chi _{1}^{0} \\ 
\chi _{2}^{-} \\ 
\frac{1}{\sqrt{2}}(\upsilon _{\chi }+\xi _{\chi }\pm i\zeta _{\chi }) \\ 
\end{pmatrix}%
:(3,-1/3),\hspace{1cm} \notag \\
\rho & =%
\begin{pmatrix}
\rho _{1}^{+} \\ 
\frac{1}{\sqrt{2}}(\upsilon _{\rho }+\xi _{\rho }\pm i\zeta _{\rho }) \\ 
\rho _{3}^{+} \\ 
\end{pmatrix}%
:(3,2/3),  \notag \\
\eta & =%
\begin{pmatrix}
\frac{1}{\sqrt{2}}(\upsilon _{\eta }+\xi _{\eta }\pm i\zeta _{\eta }) \\ 
\eta _{2}^{-} \\ 
\eta _{3}^{0}%
\end{pmatrix}%
:(3,-1/3).   \label{SU3Lscalartriplets}
\end{align}
We extend the scalar sector of the 3-3-1 model with right handed Majorana
neutrinos by adding the following $SU(3)_{L}$ scalar singlets: 
\begin{align}
\phi & :(1,0),\hspace{0.7cm}\sigma \sim (1,0),\hspace{0.7cm}\xi _{n}:(1,0),%
\hspace{0.7cm}n=1,2,  \notag \\
\tau _{j}& :(1,0),\hspace{0.7cm}\Xi _{j}:(1,0),\hspace{0.7cm}S_{j}:(1,0),%
\hspace{0.7cm}j=1,2,3,  \notag \\
\Phi _{j}& :(1,0),\hspace{0.7cm}\Omega _{j}:(1,0),\hspace{0.7cm}\Theta
_{j}:(1,0),\hspace{0.7cm}j=1,2,3.  \label{SU3Lscalarsinglets}
\end{align}

The scalar fields are assigned to different singlet and triplet representations of the $\Delta \left( 27\right) $ discrete group, as follows:
\begin{eqnarray}
\eta &\sim &\mathbf{\mathbf{1}_{\mathbf{1,0}}},\hspace{0.9cm}\rho \sim 
\mathbf{1}_{\mathbf{2,0}},\hspace{0.9cm}\chi \sim \mathbf{1}_{\mathbf{0,0}},%
\hspace{0.9cm}\sigma \sim \mathbf{1}_{\mathbf{0,0}},\notag \\
\phi &\sim &\mathbf{\mathbf{1}_{\mathbf{0,0}}},\hspace{1cm}\tau _{1}\sim 
\mathbf{1}_{\mathbf{0,0}},\hspace{1cm}\tau _{2}\sim \mathbf{1}_{\mathbf{0,2}%
},\notag \\
\tau _{3} &\sim &\mathbf{1}_{\mathbf{0,2}},\hspace{1cm}\xi _{1}\sim \mathbf{1%
}_{\mathbf{0,0}},\hspace{1cm}\xi _{2}\sim \mathbf{1}_{\mathbf{0,0}},\\
S &\sim &\mathbf{3},\hspace{0.7cm}\Xi \sim \mathbf{3},\hspace{0.7cm}\Phi
\sim \mathbf{3},\hspace{0.7cm}\Omega \sim \mathbf{3},\hspace{0.7cm}\Theta
\sim \mathbf{3}\notag.
\end{eqnarray}

The $Z_{4}\otimes Z_{8}\otimes Z_{14}$ assignments of the scalar fields are: 
\begin{eqnarray}
\eta &\sim &\left( 1,1,1\right) ,\hspace{0.5cm}\rho \sim \left( 1,1,1\right)
,\hspace{0.5cm}\chi \sim \left( i,1,e^{-\frac{i\pi }{7}}\right) ,  \notag \\
\sigma &\sim &\left( 1,1,e^{-\frac{i\pi }{7}}\right) ,\hspace{0.5cm}\phi
\sim \left( \mathbf{-}i,-1,-1\right) ,\hspace{0.5cm}  \notag \\
\tau _{1} &\sim &\left( \mathbf{-}1,-i,e^{\frac{3i\pi }{7}}\right) ,\hspace{%
0.5cm}\tau _{2}\sim \left( 1,e^{-\frac{i\pi }{4}},e^{\frac{4i\pi }{7}%
}\right) ,  \notag \\
\tau _{3} &\sim &\left( 1,-i,e^{\frac{5i\pi }{7}}\right) ,\hspace{0.5cm}\xi
_{1}\sim \left( 1,e^{-\frac{i\pi }{4}},e^{\frac{i\pi }{7}}\right) ,\hspace{%
0.5cm}\notag \\
\xi _{2} &\sim &\left( \mathbf{-}i,1,e^{\frac{2i\pi }{7}}\right) ,\hspace{%
0.5cm}S\sim \left( 1,e^{-\frac{i\pi }{4}},1\right) ,\hspace{0.5cm}  \notag \\
\Xi &\sim &\left( 1,1,e^{-\frac{2i\pi }{7}}\right) ,\hspace{0.5cm}\Phi \sim
\left( 1,1,e^{\frac{i\pi }{7}}\right) ,  \notag \\
\Omega &\sim &\left( \mathbf{-}1,1,1\right) ,\hspace{0.5cm}\Theta \sim
\left( \mathbf{-}1,1,e^{-\frac{i\pi }{7}}\right) .
\end{eqnarray}

Regarding leptons, we group left handed leptons and right handed Majorana
neutrinos in $\Delta \left( 27\right) $ triplets, whereas right handed
charged leptons are assigned as $\Delta \left( 27\right) $ triplets, as
follows: 
\begin{eqnarray}
L_{L} &\sim &\mathbf{3},\hspace{1cm}e_{R}\sim \mathbf{\mathbf{1}_{\mathbf{1,0%
}}},\hspace{1cm}\mu _{R}\sim \mathbf{\mathbf{1}_{2\mathbf{,0}}},  \notag \\
\tau _{R} &\sim &\mathbf{\mathbf{1}_{0\mathbf{,0}}},\hspace{1cm}N_{R}\sim 
\mathbf{3}.  \label{leptonassignments}
\end{eqnarray}

The $Z_{4}\otimes Z_{8}\otimes Z_{14}$ assignments for leptons are:
\begin{eqnarray}
L_{L} &\sim &\left( i,1,1\right) ,\hspace{0.3cm}e_{R}\sim \left( i,e^{\frac{%
i\pi }{4}},-1\right) ,\hspace{0.3cm}\mu _{R}\sim \left( i,e^{\frac{i\pi }{4}%
},e^{\frac{4i\pi }{7}}\right) ,  \notag \\
\tau _{R} &\sim &\left( i,e^{\frac{i\pi }{4}},e^{\frac{2i\pi }{7}}\right) ,%
\hspace{0.3cm}N_{R}\sim \left( 1,1,e^{\frac{i\pi }{7}}\right) .
\end{eqnarray}

Regarding quarks, we assign quark fields into different singlet
representations of the $\Delta \left( 27\right) $ discrete group, as
follows: 
\begin{eqnarray}
Q_{L}^{1} &\sim &\mathbf{1}_{\mathbf{0,0}},\hspace{1cm}Q_{L}^{2}\sim \mathbf{%
1}_{\mathbf{0,0}},\hspace{1cm}Q_{L}^{3}\sim \mathbf{1}_{\mathbf{0,0}}, 
\notag \\
U_{R}^{1} &\sim &\mathbf{1}_{\mathbf{2,0}},\hspace{1cm}U_{R}^{2}\sim \mathbf{%
1}_{\mathbf{2,0}},\hspace{1cm}U_{R}^{3}\sim \mathbf{1}_{\mathbf{2,0}}, 
\notag \\
D_{R}^{1} &\sim &\mathbf{1}_{\mathbf{1,0}},\hspace{1cm}D_{R}^{2}\sim \mathbf{%
1}_{\mathbf{1,1}},\hspace{1cm}D_{R}^{3}\sim \mathbf{1}_{\mathbf{1,0}}, 
\notag \\
T_{R} &\sim &\mathbf{1}_{\mathbf{0,0}},\hspace{1cm}J_{R}^{1}\sim \mathbf{1}_{%
\mathbf{0,0}},\hspace{1cm}J_{R}^{2}\sim \mathbf{1}_{\mathbf{0,0}}.
\label{Quarkassignments}
\end{eqnarray}

The $Z_{4}\otimes Z_{8}\otimes Z_{14}$ assignments for quarks are: 
\begin{eqnarray}
Q_{L}^{1} &\sim &\left( 1,-i,1\right) ,\hspace{0.3cm}Q_{L}^{2}\sim \left(
1,e^{-\frac{i\pi }{4}},1\right) ,\hspace{0.3cm}Q_{L}^{3}\sim \left(
1,1,1\right),\notag \\
U_{R}^{1} &\sim &\left( i,i,1\right) ,\hspace{0.3cm}U_{R}^{2}\sim \left( 
\mathbf{-}1,e^{\frac{i\pi }{4}},1\right) ,\hspace{0.3cm}U_{R}^{3}\sim \left(
1,1,1\right),\notag \\
D_{R}^{1} &\sim &\left( i,i,e^{-\frac{i\pi }{7}}\right) ,\hspace{0.3cm}%
D_{R}^{2}\sim \left( 1,1,1\right) ,\hspace{0.3cm}D_{R}^{3}\sim \left(
i,1,1\right),\notag \\
T_{R} &\sim &\left( -i,1,e^{\frac{i\pi }{7}}\right) ,\notag \\
J_{R}^{1} &\sim &\left( i,-i,e^{-\frac{i\pi }{7}}\right) ,\hspace{0.3cm}%
J_{R}^{2}\sim \left( i,e^{-\frac{i\pi }{4}},e^{-\frac{i\pi }{7}}\right) .
\end{eqnarray}
Here the dimensions of the $\Delta \left( 27\right) $ irreducible
representations are specified by the numbers in boldface. As regards the lepton sector, we recall that the left and right-handed leptons are
grouped into $\Delta \left( 27\right) $ triplet and $\Delta \left( 27\right) 
$ singlet irreducible representations, respectively, whe-reas the
right-handed Majorana neutrinos are unified into a $\Delta \left( 27\right) $
triplet. Regarding the quark sector, we assign the quarks fields into $%
\Delta \left( 27\right) $ singlet representations. Specifically, we assign
the left-handed $SU(3)_{L}$ quark triplets and right-handed exotic quarks as 
$\Delta \left( 27\right) $ trivial singlets, whereas the right-handed SM
quarks are assigned as $\Delta \left( 27\right) $ nontrivial singlets.
Furthermore, it is worth mentioning that the $SU(3)_{L}$ scalar triplets are
assigned to one$\ \Delta \left( 27\right) $ trivial and two $\Delta \left(
27\right) $ nontrivial singlet representations, whereas the $SU(3)_{L}$
scalar singlets are accommodated into five $\Delta \left( 27\right) $
triplets, six $\Delta \left( 27\right) $ trivial singlets and four $\Delta
\left( 27\right) $ nontrivial singlets. Out of the five $SU(3)_{L}$ scalar
singlets $\Delta \left( 27\right) $ triplets, only one is charged under the $%
Z_{8}$ discrete symmetry whereas the remaining are $Z_{8}$ neutral. As we
will see in the following, the $Z_{8}$ discrete symmetry separates the $%
\Delta \left( 27\right) $ scalar triplets participating in the charged
lepton Yukawa interactions from those one appearing in the neutrino Yukawa
terms. Furthermore, as regards the $Z_{8}$ neutral $\Delta \left(
27\right) $ scalar triplets participating in the neutrino Yukawa
interactions, it is worth mentioning that they are distinguished by their $%
Z_{4}$ charges. Those $Z_{8}$ neutral $\Delta \left( 27\right) $ scalar
triplets transforming trivially under the $Z_{4}$ symmetry, contribute to
the right-handed Majorana neutrino mass matrix, whereas the remaining $Z_{8}$
neutral $\Delta \left( 27\right) $ triplet scalar fields are $Z_{4}$ charged
and give rise to the Dirac neutrino mass matrix.

With the above particle content, the following Yukawa terms for the quark
and lepton sectors arise: 
\begin{eqnarray}
-\mathcal{L}_{Y}^{\left( Q\right) } &=&y_{11}^{\left( U\right) }\overline{Q}%
_{L}^{1}\rho ^{\ast }U_{R}^{1}\frac{\phi \sigma ^{7}}{\Lambda ^{8}}%
+y_{22}^{\left( U\right) }\overline{Q}_{L}^{2}\rho ^{\ast }U_{R}^{2}\frac{%
\tau _{1}\sigma ^{3}}{\Lambda ^{4}}  \notag \\
&&+y_{23}^{\left( U\right) }\overline{Q}_{L}^{2}\rho ^{\ast }U_{R}^{3}\frac{%
\xi _{1}\sigma }{\Lambda ^{2}}+y_{13}^{\left( U\right) }\overline{Q}%
_{L}^{1}\rho ^{\ast }U_{R}^{3}\frac{\xi _{1}^{2}\sigma ^{2}}{\Lambda ^{4}} 
\notag \\
&&+y_{33}^{\left( U\right) }\overline{Q}_{L}^{3}\eta
U_{R}^{3}+y_{11}^{\left( D\right) }\overline{Q}_{L}^{1}\eta ^{\ast }D_{R}^{1}%
\frac{\phi \sigma ^{6}}{\Lambda ^{7}}  \label{Lyq} \\
&&+y_{22}^{\left( D\right) }\overline{Q}_{L}^{2}\eta ^{\ast }D_{R}^{2}\frac{%
\tau _{2}\sigma ^{4}}{\Lambda ^{5}}+y_{12}^{\left( D\right) }\overline{Q}%
_{L}^{1}\eta ^{\ast }D_{R}^{2}\frac{\tau _{3}\sigma ^{5}}{\Lambda ^{6}} 
\notag \\
&&+y_{33}^{\left( D\right) }\overline{Q}_{L}^{3}\rho D_{R}^{3}\frac{\xi
_{2}\sigma ^{2}}{\Lambda ^{3}}+y_{1}^{\left( J\right) }\overline{Q}%
_{L}^{1}\chi ^{\ast }J_{R}^{1}  \notag \\
&&+y_{2}^{\left( J\right) }\overline{Q}_{L}^{2}\chi ^{\ast
}J_{R}^{2}+y^{\left( T\right) }\overline{Q}_{L}^{3}\chi T_{R}+H.c,  \notag
\end{eqnarray}%
\begin{eqnarray}
-\mathcal{L}_{Y}^{\left( L\right) } &=&h_{\rho e}^{\left( L\right) }\left( 
\overline{L}_{L}\rho S\right) _{\mathbf{\mathbf{1}_{0\mathbf{,0}}}}e_{R}%
\frac{\sigma ^{7}}{\Lambda ^{8}}+h_{\rho \mu }^{\left( L\right) }\left( 
\overline{L}_{L}\rho S\right) _{\mathbf{1}_{\mathbf{1,0}}}\mu _{R}\frac{%
\sigma ^{4}}{\Lambda ^{5}}  \notag \\
&&+h_{\rho \tau }^{\left( L\right) }\left( \overline{L}_{L}\rho S\right) _{%
\mathbf{\mathbf{1}_{\mathbf{2,0}}}}\tau _{R}\frac{\sigma ^{2}}{\Lambda ^{3}}%
+h_{\chi }^{\left( L\right) }\left( \overline{L}_{L}\chi N_{R}\right) _{%
\mathbf{\mathbf{1}_{0\mathbf{,0}}}}  \notag \\
&&+\frac{h_{1N}}{2}\left( \overline{N}_{R}N_{R}^{C}\right) _{\mathbf{3}%
_{S_{1}}}\Xi ^{\ast }+\frac{h_{2N}}{2}\left( \overline{N}_{R}N_{R}^{C}%
\right) _{\mathbf{3}_{S_{1}}}\Phi ^{\ast }\frac{\sigma }{\Lambda }  \notag \\
&&+h_{3N}\left( \overline{N}_{R}N_{R}^{C}\right) _{\mathbf{3}_{S_{2}}}\Xi
^{\ast }+h_{4N}\left( \overline{N}_{R}N_{R}^{C}\right) _{\mathbf{3}%
_{S_{2}}}\Phi ^{\ast }\frac{\sigma }{\Lambda }  \notag \\
&&+h_{\rho }^{\left( 1\right) }\varepsilon _{abc}\left( \overline{L}%
_{L}^{a}\left( L_{L}^{C}\right) ^{b}\right) _{\mathbf{3}_{S_{2}}}\left( \rho
^{\ast }\right) ^{c}\frac{\Omega ^{\ast }}{\Lambda }  \label{Lyl} \\
&&+h_{\rho }^{\left( 2\right) }\varepsilon _{abc}\left( \overline{L}%
_{L}^{a}\left( L_{L}^{C}\right) ^{b}\right) _{\mathbf{3}_{S_{2}}}\left( \rho
^{\ast }\right) ^{c}\frac{\Theta ^{\ast }\sigma }{\Lambda ^{2}}+H.c,  \notag
\end{eqnarray}%
where $y_{ij}^{\left( U,D\right) }$ ($i,j=1,2,3$), $y^{\left( T\right) }$, $%
y_{m}^{\left( J\right) }$, $\ h_{\rho }^{\left( m\right) }$\ ($m=1,2$), $%
h_{sN}$ ($s=1,2,3,4$), $h_{\chi }^{\left( L\right) }$,$\ h_{\rho e}^{\left(
L\right) }$, $h_{\rho \mu }^{\left( L\right) }$ and $h_{\rho \tau }^{\left(
L\right) }$ are $\mathcal{O}(1)$ dimensionless couplings. We assume that all
of these dimensionless couplings are real, except for $y_{13}^{\left(
U\right) }$, $h_{\rho \mu }^{\left( L\right) }$ and $h_{\rho \tau }^{\left(
L\right) }$, taken to be complex. In the following we provide a justification for the aforementioned assumption. As shown in Sect. \ref{leptonmassesandmixing}, having $h_{\rho \mu }^{\left( L\right) }$ and $h_{\rho \tau }^{\left(L\right) }$ complex is required to yield leptonic mixing angles consistent with the current neutrino oscillation experimental data. Furthermore, as shown in Sect. \ref{quarkmassesandmixing}, the quark assignments under the different group factors of our model will give rise to SM quark mass texture where the Cabbibo mixing arise from the down type quark sector, whereas the up type quark sector contributes to the remaining mixing angles. As indicated by the current low energy quark flavor data encoded in the Standard parametrization of the quark mixing matrix, the complex phase responsible for CP violation in the quark sector is associated with the quark mixing angle in the $1$-$3$ plane. Consequently, in order to reproduce the experimental values of quark mixing angles and CP violating phase, $y_{13}^{\left(U\right) }$ is required to be complex. 

\quad An explanation of the role of each discrete group factor of our model
is provided in the following. The $\Delta \left(27\right) $, $Z_{4}$ and $%
Z_{8}$ discrete groups are crucial for reducing the number of model
parameters, thus increasing the predictivity of our model and giving rise to
predictive and viable textures for the fermion sector, consistent with the
observed pattern of fermion masses and mixings, as will be shown later in
Sects. \ref{leptonmassesandmixing} and \ref{quarkmassesandmixing}. The $%
Z_{4}$ and $Z_{14}$ symmetries reduce the number of parameters in the
neutrino sector. Besides that, the $Z_{4}$ and $Z_{8}$ discrete group
determine the allowed entries of the SM quark mass matrices. As a result of
the $Z_{4}\otimes Z_{8}$ charge assignments for the SM quark sector given by
Eq. (\ref{Quarkassignments}), the Cabbibo mixing will arise from the down
type quark sector, whereas the up sector will contribute to the remaining
mixing angles. Furthermore, thanks to the $\Delta \left(27\right) $
discrete symmetry, SM quarks do not mix with the exotic ones. This arises
from the fact that the right-handed SM and exotic quarks are assigned as
nontrivial and trivial $\Delta \left( 27\right) $ singlets, respectively.
The $Z_{14}$ symmetry give rises to the hierarchical structure of the
charged fermion mass matrices that yields the observed charged fermion mass
and quark mixing pattern. Let us note that the five dimensional Yukawa
operators $\frac{1}{\Lambda }\left( \overline{L}_{L}\rho S\right) _{\mathbf{%
\mathbf{1}_{0\mathbf{,0}}}}e_{R}$, $\frac{1}{\Lambda }\left( \overline{L}%
_{L}\rho S\right) _{\mathbf{1}_{\mathbf{1,0}}}\mu _{R}$ and $\frac{1}{%
\Lambda }\left( \overline{L}_{L}\rho S\right) _{\mathbf{1}_{\mathbf{2,0}%
}}\tau _{R}$ are invariant under the $\Delta \left( 27\right) $ family
symmetry but do not respect the $Z_{14}$ symmetry, as the right-handed
charged leptons transform nontrivially under the $Z_{14}$ cyclic group. Let
us note that the $Z_{14}$ symmetry is the smallest lowest cyclic symmetry,
from which charged lepton Yukawa term of dimension 12 can be built, by
inserting $\frac{\sigma ^{7}}{\Lambda ^{7}}$ on the $\frac{1}{\Lambda }%
\left( \overline{L}_{L}\rho S\right) _{\mathbf{\mathbf{1}_{0\mathbf{,0}}}%
}e_{R}$ operator. It is noteworthy that the small value of the electron mass
can naturally arise from the aforementioned charged lepton Yukawa term of
dimension 12.

Furthermore, since the breaking of the $\Delta \left( 27\right) \otimes
Z_{4}\otimes Z_{8}\otimes Z_{14}$ discrete group gives rise to the charged
fermion mass and quark mixing pattern, we set the VEVs of the $SU(3)_{L}$ singlet scalar fields $\phi$, $\xi _{n}$ ($n=1,2$), $\tau _{j}$, $S_{j}$ $(j=1,2,3)$ and $\sigma $, with respect to the Wolfenstein parameter $\lambda =0.225$ and the model cutoff $\Lambda$, as follows: 
\begin{equation}
v_{S}\sim v_{\phi }\sim v_{\tau _{1}}\sim v_{\tau _{2}}\sim v_{\tau
_{3}}\sim v_{\xi _{1}}\sim v_{\xi _{2}}\sim v_{\sigma }\sim \Lambda
_{int}=\lambda \Lambda ,  \label{VEV}
\end{equation}
Let us note that the $SU(3)_{L}$ singlet scalar fields $\phi$, $\xi _{n}$ ($n=1,2$), $\tau _{j}$, $S_{j}$, $\Omega _{j}$, $\Theta _{j}$ ($j=1,2,3$) and $\sigma $ having the VEVs of the same order of magnitude are the ones that appear in the SM charged fermion Yukawa terms, thus playing an important role in generating the SM charged fermion masses and quark mixing angles.

As we will explain in the following, we are going to implement a double seesaw mechanism for the generation of the light active neutrino masses. To implement a double seesaw mechanism, we need very heavy right handed Majorana neutrinos, which implies that the $SU(3)_{L}$ singlet scalars should acquire very large vacuum expectation values. In addition, in order to simplify our analysis of the scalar potential for the $\Delta(27)$ scalar triplets, we need that the $\Delta(27)$ scalar triplets $\Xi$ and $\Phi$ contributing to the right handed Majorana neutrino masses should acquire much lower VEVs than the $\Delta(27)$ scalar triplet $S$ that gives rise to the charged lepton masses. That hierarchy in their VEVs will allow to neglect the mixings between these fields as follows from the method of recursive expansion of Ref. \cite{Grimus:2000vj} and to treat their scalar potentials independently. Because of these reasons, we assume that the VEVs of $SU(3)_{L}$ singlet scalar fields $\Xi _{j}$, $\Phi _{j}$ ($j=1,2,3$) satisfy the following hierarchy:
\begin{equation}
v_{\chi }\ll v_{\Xi }\sim v_{\Phi }\ll \Lambda _{int}.
\end{equation}
Furthermore, implementing a double seesaw mechanism also requires that the $\Delta(27)$ scalar triplets $\Omega$ and $\Theta$ contributing to the Dirac neutrino Yukawa terms, should acquire VEVs much lower than the electroweak symmetry breaking scale $v=246$ GeV. Consequently, the scalar fields of our model obey the following hierarchy: 
\begin{equation}
v_{\Omega }\sim v_{\Theta }\ll v_{\rho }\sim v_{\eta }\sim v\ll v_{\chi }\ll
v_{\Xi }\sim v_{\Phi }\ll \Lambda _{int}.  \label{VEVsize}
\end{equation}%
Thus, the $SU(3)_{L}$ scalar singlets presented in the right-handed Majorana
neutrino Yukawa interactions, acquire very large vacuum expectation values,
which implies that the Majorana neutrinos acquire very large masses, hence
allowing to implement a double seesaw mechanism to generate the light active
neutrino masses. Consequently, the neutrino spectrum is composed of very
light active neutrinos as well as heavy and very heavy sterile neutrinos.

\quad In summary, for the reasons mentioned above and considering a very high model cutoff $\Lambda\gg v_{\chi}$, we set the vacuum expectation values (VEVs) of the $SU(3)_{L}$
scalar singlets at a very high energy, much larger than $v_{\chi }\approx \mathcal{O}(1)$ TeV, with the exception
of the VEVs of $\Omega _{j}$ and $\Theta _{j}$ ($j=1,2,3$), taken to be much
smaller than the electroweak symmetry breaking scale $v=246$ GeV. It is
noteworthy the $SU(3)_{C}\otimes SU\left( 3\right) _{L}\otimes U\left(
1\right) _{X}\otimes \Delta \left( 27\right) \otimes Z_{4}\otimes
Z_{8}\otimes Z_{14}$ symmetry is broken down to $SU(3)_{C}\otimes
SU(3)_{L}\otimes U(1)_{X}$, at the scale $\Lambda _{int}$, by the vacuum
expectation values of the $SU(3)_{L}$ singlet scalar fields $\phi $, $\xi
_{n}$ ($n=1,2$), $\tau _{j}$, $S_{j}$, ($j=1,2,3$) and $\sigma $.

\quad In the following we comment on the possible VEV patterns for the $\Delta(27)$
scalar triplets $S$, $\Xi$, $\Phi$, $\Omega$, and $\Theta$. Since the VEVs of the $\Delta(27)$
scalar triplets satisfy the following hierarchy: $v_{\Omega }\sim v_{\Theta }\ll v_{\Xi }\sim v_{\Phi }\ll v_S$  the mixing angles
of $S$ and $\Xi$ with $\Phi$, $\Omega$, and $v_S$ are very small since they are suppressed by the
ratios of their VEVs, which is a consequence of the method of recursive
expansion proposed in Ref. \cite{Grimus:2000vj}. Thus, the scalar potential for the $\Delta(27)$ scalar triplet $S$ can be treated independently from the scalar potentials for the two sets of $\Delta(27)$ scalar triplets $\Xi$, $\Phi$, and $\Omega$ and $\Theta$. Furthermore, because of the reason mentioned above, one can treat the scalar potential for $\Xi$, $\Phi$ independently from the one that involves  $\Omega$ and $\Theta$. As shown in detail in Appendix \ref{ApB}, the following VEV patterns for the $\Delta (27)$ scalar triplets are consistent with the scalar potential minimization equations for a large region of parameter space: 
\begin{eqnarray}
\left\langle S\right\rangle &=&\frac{v_{S}}{\sqrt{3}}\left( 1,1,1\right) ,%
\hspace{1cm}\left\langle \Xi \right\rangle =v_{\Xi }\left( 1,0,0\right) , 
\notag \\
\left\langle \Phi \right\rangle &=&v_{\Phi }\left( 0,0,1\right) ,\hspace{1cm}%
\left\langle \Omega \right\rangle =v_{\Omega }\left( 1,0,0\right) ,  \notag
\\
\left\langle \Theta \right\rangle &=&v_{\Theta }\left( 0,0,1\right) .
\label{VEVpattern}
\end{eqnarray}%

\section{Lepton masses and mixings}
\label{leptonmassesandmixing}From the lepton Yukawa terms given by Eq. (\ref{Lyl}), we find that the mass matrix for charged leptons takes the form: 
\begin{eqnarray}
M_{l} &=&R_{lL}^{\dag }P_{l}diag\left( m_{e},m_{\mu },m_{\tau }\right) ,%
\hspace{0.3cm}R_{lL}=\frac{1}{\sqrt{3}}\left( 
\begin{array}{ccc}
1 & 1 & 1 \\ 
1 & \omega & \omega ^{2} \\ 
1 & \omega ^{2} & \omega%
\end{array}%
\right) ,  \notag \\
P_{l} &=&\left( 
\begin{array}{ccc}
1 & 0 & 0 \\ 
0 & e^{i\alpha } & 0 \\ 
0 & 0 & e^{i\beta }%
\end{array}%
\right) ,\hspace{0.5cm}\omega =e^{\frac{2\pi i}{3}},  \label{Ml}
\end{eqnarray}%
$\alpha $ and $\beta$ being the complex phases of $h_{\rho \mu }^{\left(
L\right) }$ and $h_{\rho \tau }^{\left( L\right) }$, respectively, and the
charged lepton masses given by: 
\begin{equation}
m_{e}=a_{1}^{\left( l\right) }\lambda ^{8}\frac{v}{\sqrt{2}},\hspace{0.5cm}%
m_{\mu }=a_{2}^{\left( l\right) }\lambda ^{5}\frac{v}{\sqrt{2}},\hspace{0.5cm%
}m_{\tau }=a_{3}^{\left( l\right) }\lambda ^{3}\frac{v}{\sqrt{2}}.
\label{leptonmasses}
\end{equation}%
$\lambda =0.225$ is one of the Wolfenstein parameters, $v=246$ GeV the
electroweak symmetry breaking scale, and $a_{i}^{\left( l\right) }$ ($i=1,2,3$%
) $\mathcal{O}(1)$ dimensionless parameters. Let us note that the charged
lepton masses are connected with the scale of electroweak symmetry breaking,
through their power dependence on the Wolfenstein parameter $\lambda =0.225$%
, with $\mathcal{O}(1)$ coefficients.

\quad Regarding the neutrino sector, we see that the neutrino mass terms
take the form: 
\begin{equation}
-\mathcal{L}_{mass}^{\left( \nu \right) }=\frac{1}{2}\left( 
\begin{array}{ccc}
\overline{\nu _{L}^{C}} & \overline{\nu _{R}} & \overline{N_{R}}%
\end{array}%
\right) M_{\nu }\left( 
\begin{array}{c}
\nu _{L} \\ 
\nu _{R}^{C} \\ 
N_{R}^{C}%
\end{array}%
\right) +H.c,  \label{Lnu}
\end{equation}%
where the $\Delta \left( 27\right) $ family symmetry constrains the neutrino
mass matrix to be of the form: 
\begin{eqnarray}
M_{\nu } &=&\left( 
\begin{array}{ccc}
0_{3\times 3} & M_{D} & 0_{3\times 3} \\ 
M_{D}^{T} & 0_{3\times 3} & M_{\chi } \\ 
0_{3\times 3} & M_{\chi }^{T} & M_{R}%
\end{array}%
\right) ,\hspace{0.4cm}  \notag \\
M_{D} &=&\frac{h_{\rho }^{\left( 1\right) }v_{\rho }v_{\Omega }}{2\Lambda }%
\left( 
\begin{array}{ccc}
0 & 1 & 0 \\ 
-1 & 0 & -a \\ 
0 & a & 0%
\end{array}%
\right) , \\
M_{\chi } &=&h_{\chi }^{\left( L\right) }\frac{v_{\chi }}{\sqrt{2}}\left( 
\begin{array}{ccc}
1 & 0 & 0 \\ 
0 & 1 & 0 \\ 
0 & 0 & 1%
\end{array}%
\right) ,\hspace{0.4cm}M_{R}=h_{1N}v_{\Xi }\left( 
\begin{array}{ccc}
1 & y & 0 \\ 
y & 0 & x \\ 
0 & x & z%
\end{array}%
\right) ,  \notag \\
x &=&\frac{h_{3N}}{h_{1N}},\hspace{0.3cm}y=\frac{h_{4N}v_{\Phi }}{%
h_{1N}v_{\Xi }},\hspace{0.3cm}z=\frac{h_{2N}v_{\Phi }v_{\sigma }}{%
h_{1N}v_{\Xi }\Lambda },\hspace{0.3cm}a=\frac{h_{\rho }^{\left( 2\right)
}v_{\Theta }v_{\sigma }}{h_{\rho }^{\left( 1\right) }v_{\Omega }\Lambda }. 
\notag
\end{eqnarray}%
As the $SU(3)_{L}$ scalar singlets presented in the right-handed Majorana
neutrino Yukawa interactions, acquire very large vacuum expectation values,
the Majorana neutrinos are very heavy, thus giving rise to a double seesaw
mechanism that generates small masses for the active neutrinos.

The neutrino mass matrix is diagonalized by a rotation matrix, which is
approximately given by \cite{Catano:2012kw}: 
\begin{equation}
\mathbb{U}=%
\begin{pmatrix}
R_{\nu } & B_{2}U_{\chi } & 0 \\ 
-B_{2}^{\dagger }R_{\nu } & U_{\chi } & B_{1}U_{R} \\ 
0 & B_{1}^{\dagger }U_{\chi } & U_{R}%
\end{pmatrix}%
,  \label{U}
\end{equation}%
with 
\begin{equation}
B_{1}^{\dagger }=M_{R}^{-1}M_{\chi }^{T},\hspace{1cm}\hspace{1cm}%
B_{2}^{\dagger }=M_{D}\left( M_{\chi }^{T}\right) ^{-1}M_{R}M_{\chi }^{-1},
\label{B}
\end{equation}%
and the neutrino mass matrices for the physical states take the form: 
\begin{eqnarray}
M_{\nu }^{\left( 1\right) } &=&M_{D}\left( M_{\chi }^{T}\right)
^{-1}M_{R}M_{\chi }^{-1}M_{D}^{T},  \label{Mnu1} \\
M_{\nu }^{\left( 2\right) } &=&-M_{\chi }M_{R}^{-1}M_{\chi }^{T},\hspace{1cm}%
\hspace{1cm}  \label{Mnu2} \\
M_{\nu }^{\left( 3\right) } &=&M_{R},  \label{Mnu3}
\end{eqnarray}%
where $M_{\nu }^{\left( 1\right) }$ is the light active neutrino mass
matrix, whereas $M_{\nu }^{\left( 2\right) }$ and $M_{\nu }^{\left( 3\right)
}$ are the heavy and very heavy sterile neutrino mass matrices,
respectively. Thus, the double seesaw mechanism produces a neutrino spectrum
composed of light active neutrinos, heavy and very heavy sterile neutrinos.
Furthermore, let us note that the neutrino mass matrices $M_{\nu }^{\left(
1\right) }$, $M_{\nu }^{\left( 2\right) }$, and $M_{\nu }^{\left( 3\right) }$
are diagonalized by the rotation matrices $R_{\nu }$, $U_{R}$ and $U_{\chi }$%
, respectively \cite{Catano:2012kw}.

\quad Using Eq. (\ref{Mnu1}), we find that the light active neutrino mass
matrix takes the form: 
\begin{eqnarray}
M_{\nu }^{\left( 1\right) } &=&\frac{h_{1N}\left( h_{\rho }^{\left( 1\right)
}\right) ^{2}v_{\rho }^{2}v_{\Omega }^{2}v_{\Xi }}{2\left( h_{\chi }^{\left(
L\right) }\right) ^{2}v_{\chi }^{2}\Lambda ^{2}}\left( 
\begin{array}{ccc}
0 & 1 & 0 \\ 
-1 & 0 & -a \\ 
0 & a & 0%
\end{array}%
\right) \left( 
\begin{array}{ccc}
1 & y & 0 \\ 
y & 0 & x \\ 
0 & x & z%
\end{array}%
\right)\left( 
\begin{array}{ccc}
0 & -1 & 0 \\ 
1 & 0 & a \\ 
0 & -a & 0%
\end{array}%
\right)  \notag \\
&=&\frac{h_{1N}\left( h_{\rho }^{\left( 1\right) }\right) ^{2}v_{\rho
}^{2}v_{\Omega }^{2}v_{\Xi }}{2\left( h_{\chi }^{\left( L\right) }\right)
^{2}v_{\chi }^{2}\Lambda ^{2}}\left( 
\begin{array}{ccc}
0 & -y-ax & 0 \\ 
-y-ax & za^{2}+1 & -a\left( y+ax\right) \\ 
0 & -a\left( y+ax\right) & 0%
\end{array}%
\right) \allowbreak  \notag \\
&=&\left( 
\begin{array}{ccc}
0 & A & 0 \\ 
A & C & B \\ 
0 & B & 0%
\end{array}%
\right) .  \label{Mnu}
\end{eqnarray}

Then we find that, for the normal (NH) and inverted (IH) neutrino mass
hierarchies, the light active neutrino mass matrix is diagonalized by a
rotation matrix $R_{\nu }$, according to: 
\begin{eqnarray}
R_{\nu }^{T}M_{\nu }^{\left( 1\right) }R_{\nu } &=&\left( 
\begin{array}{ccc}
0 & 0 & 0 \\ 
0 & m_{\nu _{2}} & 0 \\ 
0 & 0 & m_{\nu _{3}}%
\end{array}%
\right) \allowbreak ,\hspace{1cm}m_{\nu _{1}}=0,  \notag \\
m_{\nu _{2,3}} &=&\frac{C}{2}\mp \frac{1}{2}\sqrt{K},\hspace{1cm}%
\mbox{for
NH},  \label{NeutrinomassNH} \\
R_{\nu } &=&\left( 
\begin{array}{ccc}
-\frac{B}{\sqrt{A^{2}+B^{2}}} & \frac{\sqrt{2}A}{\sqrt{K-C\sqrt{K}}} & \frac{%
\sqrt{2}A}{\sqrt{K+C\sqrt{K}}} \\ 
0 & \frac{1}{\sqrt{2}}\frac{C-\sqrt{K}}{\sqrt{K-C\sqrt{K}}} & \frac{1}{\sqrt{%
2}}\frac{C+\sqrt{K}}{\sqrt{K+C\sqrt{K}}} \\ 
\frac{A}{\sqrt{A^{2}+B^{2}}} & \frac{\sqrt{2}B}{\sqrt{K-C\sqrt{K}}} & \frac{%
\sqrt{2}B}{\sqrt{K+C\sqrt{K}}}%
\end{array}%
\right) .  \notag
\end{eqnarray}%
\begin{eqnarray}
R_{\nu }^{T}M_{\nu }^{\left( 1\right) }R_{\nu } &=&\left( 
\begin{array}{ccc}
m_{\nu _{1}} & 0 & 0 \\ 
0 & m_{\nu _{2}} & 0 \\ 
0 & 0 & 0%
\end{array}%
\right) \allowbreak ,\hspace{1cm}m_{\nu _{1,2}}=\frac{C}{2}\mp \frac{1}{2}%
\sqrt{K},  \notag \\
m_{\nu _{3}} &=&0,\hspace{1cm}\mbox{for
IH},  \label{NeutrinomassIH} \\
\allowbreak R_{\nu } &=&\left( 
\begin{array}{ccc}
\frac{\sqrt{2}A}{\sqrt{K-C\sqrt{K}}} & \frac{\sqrt{2}A}{\sqrt{K+C\sqrt{K}}}
& -\frac{B}{\sqrt{A^{2}+B^{2}}} \\ 
\frac{1}{\sqrt{2}}\frac{C-\sqrt{K}}{\sqrt{K-C\sqrt{K}}} & \frac{1}{\sqrt{2}}%
\frac{C+\sqrt{K}}{\sqrt{K+C\sqrt{K}}} & 0 \\ 
\frac{\sqrt{2}B}{\sqrt{K-C\sqrt{K}}} & \frac{\sqrt{2}B}{\sqrt{K+C\sqrt{K}}}
& \frac{A}{\sqrt{A^{2}+B^{2}}}%
\end{array}%
\right) ,  \notag
\end{eqnarray}%
where 
\begin{equation}
K=4A^{2}+4B^{2}+C^{2}.
\end{equation}%
Using the rotation matrices in the charged lepton sector $V_{L}$, given by
Eq. (\ref{Ml}), and in the neutrino sector $R_{\nu }$, given by Eqs. (\ref%
{NeutrinomassNH}) and (\ref{NeutrinomassIH}) for normal (NH) and inverted
(IH) neutrino mass hierarchies, respectively, we find that the
Pontecorvo-Maki-Nakagawa-Sakata (PMNS) leptonic mixing matrix takes the form:
%
\begin{widetext}
\begin{eqnarray}
U=R_{lL}^{\dag }P_{l}R_{\nu }=\left\{ 
\begin{array}{l}
\left( 
\begin{array}{ccc}
-\frac{B-Ae^{i\beta }}{\sqrt{3}\sqrt{A^{2}+B^{2}}} & \frac{2A+2Be^{i\beta
}+e^{i\alpha }\left( C-\sqrt{K}\right) }{\sqrt{6}\sqrt{K-C\sqrt{K}}} & \frac{%
2A+2Be^{i\beta }+e^{i\alpha }\left( C+\sqrt{K}\right) }{\sqrt{6}\sqrt{K+C%
\sqrt{K}}} \\ 
-\frac{B-A\omega e^{i\beta }}{\sqrt{3}\sqrt{A^{2}+B^{2}}} & \frac{%
2A+2B\omega e^{i\beta }+\omega ^{2}e^{i\alpha }\left( C-\sqrt{K}\right) }{%
\sqrt{6}\sqrt{K-C\sqrt{K}}} & \frac{2A+2B\omega e^{i\beta }+\omega
^{2}e^{i\alpha }\left( C+\sqrt{K}\right) }{\sqrt{6}\sqrt{K+C\sqrt{K}}} \\ 
-\frac{B-A\omega ^{2}e^{i\beta }}{\sqrt{3}\sqrt{A^{2}+B^{2}}} & \frac{%
2A+2B\omega ^{2}e^{i\beta }+\omega e^{i\alpha }\left( C-\sqrt{K}\right) }{%
\sqrt{6}\sqrt{K-C\sqrt{K}}} & \frac{2A+2B\omega ^{2}e^{i\beta }+\omega
e^{i\alpha }\left( C+\sqrt{K}\right) }{\sqrt{6}\sqrt{K+C\sqrt{K}}}%
\end{array}%
\right)\hspace{0.1cm}\mbox{for NH} \\ 
\\ 
\left( 
\begin{array}{ccc}
\frac{2A+2Be^{i\beta }+e^{i\alpha }\left( C-\sqrt{K}\right) }{\sqrt{6}\sqrt{%
K-C\sqrt{K}}} & \frac{2A+2Be^{i\beta }+e^{i\alpha }\left( C+\sqrt{K}\right) 
}{\sqrt{6}\sqrt{K+C\sqrt{K}}} & -\frac{B-Ae^{i\beta }}{\sqrt{3}\sqrt{%
A^{2}+B^{2}}} \\ 
\frac{2A+2B\omega e^{i\beta }+\omega ^{2}e^{i\alpha }\left( C-\sqrt{K}%
\right) }{\sqrt{6}\sqrt{K-C\sqrt{K}}} & \frac{2A+2B\omega e^{i\beta }+\omega
^{2}e^{i\alpha }\left( C+\sqrt{K}\right) }{\sqrt{6}\sqrt{K+C\sqrt{K}}} & -%
\frac{B-A\omega e^{i\beta }}{\sqrt{3}\sqrt{A^{2}+B^{2}}} \\ 
\frac{2A+2B\omega ^{2}e^{i\beta }+\omega e^{i\alpha }\left( C-\sqrt{K}%
\right) }{\sqrt{6}\sqrt{K-C\sqrt{K}}} & \frac{2A+2B\omega ^{2}e^{i\beta
}+\omega e^{i\alpha }\left( C+\sqrt{K}\right) }{\sqrt{6}\sqrt{K+C\sqrt{K}}}
& -\frac{B-A\omega ^{2}e^{i\beta }}{\sqrt{3}\sqrt{A^{2}+B^{2}}}%
\end{array}%
\right)\hspace{0.1cm}\mbox{for IH}.%
\end{array}
\right.\label{PMNS}
\end{eqnarray}%
\end{widetext}

Let us note that, according to Eqs. (\ref{Ml}), (\ref{NeutrinomassNH}) and (\ref{NeutrinomassIH}), the lepton sector of our model is described by 8
effective free parameters that are fitted to reproduce the experimental
values of the 8 physical observables in the lepton sector, i.e., the three
charged lepton masses, the two neutrino mass squared splittings and the
three leptonic mixing angles. Despite this parametric freedom, we found that
the normal hierarchy scenario of our model leads to a large value of the
reactor mixing angle, not consistent with the experimental data on neutrino
oscillations. On the contrary, for the case of inverted hierarchy, as we
will see in the following, our obtained physical parameters in the lepton
sector are in excellent agreement with the experimental data. We fit the
parameters $A$, $B$, $C$, $\alpha $ and $\beta $ to reproduce the
experimental values of the neutrino mass squared splittings and three
leptonic mixing angles. By varying the parameters $A$, $B$, $C$, $\alpha $
and $\beta $, we find the following best fit result: 
\begin{eqnarray}
m_{\nu _{2}} &=&\sqrt{\Delta m_{21}^{2}+\Delta m_{13}^{2}}\approx 50\,%
\mbox{meV},\ \ \ \ \   \label{IH} \\
m_{\nu _{1}} &=&\sqrt{\Delta m_{13}^{2}}\approx 49\,\mbox{meV},\ \ \ \
\alpha \simeq -60^{\circ },\ \ \ \ \beta \simeq -165^{\circ },  \notag \\
\sin ^{2}\theta _{12} &=&0.323,\ \ \ \sin ^{2}\theta _{23}=0.573,\ \ \ \
\,\sin ^{2}\theta _{13}=0.0240,\   \notag \\
\delta &\simeq &34^{\circ },\hspace{0.2cm}J\simeq 1.96\times 10^{-2},\hspace{%
0.2cm}A\simeq -2.94\times 10^{-2}\,\mbox{meV},  \notag \\
B &\simeq &3.92\times 10^{-2}\,\mbox{meV},\hspace{0.5cm}C\simeq 7.76\times
10^{-4}\,\mbox{meV}\,,  \notag
\end{eqnarray}%
%
%
%
%
%
%
%
%
%
%
\begin{table*}[t]
\begin{tabular}{|c|c|c|c|c|c|}
\hline
Parameter & $\Delta m_{21}^{2}$($10^{-5}$eV$^{2}$) & $\Delta m_{13}^{2}$($%
10^{-3}$eV$^{2}$) & $\left( \sin ^{2}\theta _{12}\right) _{\exp }$ & $\left(
\sin ^{2}\theta _{23}\right) _{\exp }$ & $\left( \sin ^{2}\theta
_{13}\right) _{\exp }$ \\ \hline
Best fit & $7.60$ & $2.38$ & $0.323$ & $0.573$ & $0.0240$ \\ \hline
$1\sigma $ range & $7.42-7.79$ & $2.32-2.43$ & $0.307-0.339$ & $0.530-0.598$
& $0.0221-0.0259$ \\ \hline
$2\sigma $ range & $7.26-7.99$ & $2.26-2.48$ & $0.292-0.357$ & $0.432-0.621$
& $0.0202-0.0278$ \\ \hline
$3\sigma $ range & $7.11-8.11$ & $2.20-2.54$ & $0.278-0.375$ & $0.403-0.640$
& $0.0183-0.0297$ \\ \hline
\end{tabular}%
\caption{Experimental ranges of neutrino squared mass differences and
leptonic mixing angles, from Ref. \protect\cite{Forero:2014bxa}, for the
case of inverted neutrino mass spectrum.}
\label{NeutrinoobservablesIH}
\end{table*}
Comparing Eq. (\ref{IH}) with Table \ref{NeutrinoobservablesIH} we see that
the leptonic mixing parameters $\sin ^{2}\theta _{12}$, $\sin ^{2}\theta
_{13}$ and $\sin ^{2}\theta _{23}$ and the neutrino mass squared splittings are
in excellent agreement with the experimental data. We found a leptonic Dirac
CP violating phase close to $34^{\circ }$ and a Jarlskog invariant of about $%
10^{-2}$.

Now we compute the effective Majorana neutrino mass parameter, which is
proportional to the neutrinoless double beta ($0\nu \beta \beta $) decay
amplitude. The effective Majorana neutrino mass parameter is given by: 
\begin{equation}
m_{\beta \beta }=\left\vert \sum_{j}U_{ek}^{2}m_{\nu _{k}}\right\vert ,
\label{mee}
\end{equation}%
being $U_{ej}^{2}$ the PMNS mixing matrix elements and $m_{\nu _{k}}$ the
Majorana neutrino masses.

From Eqs. (\ref{PMNS}), (\ref{IH}) and (\ref{mee}), we obtain the following
value for the effective Majorana neutrino mass parameter, for the case of an inverted mass hierarchy: 
\begin{equation}
m_{\beta \beta }\approx 22\ \mbox{meV}.  \label{meevalues}
\end{equation}%
Then we get a value for the Majorana neutrino mass parameter within the
declared reach of the next-generation bolometric CUORE experiment \cite{Alessandria:2011rc} or, more realistically, of the next-to-next-generation
ton-scale $0\nu \beta \beta $-decay experiments. It is worth mentioning
that the upper limit of the Majorana neutrino mass parameter is $m_{\beta
\beta }\leq 160$ meV, which corresponds to $T_{1/2}^{0\nu \beta \beta
}(^{136}\mathrm{Xe})\geq 1.6\times 10^{25}$ yr at 90\% C.L, as follows from
the EXO-200 experiment \cite{Auger:2012ar}. It is expected an improvement of
this bound within a not too far future. The GERDA \textquotedblleft
phase-II\textquotedblright experiment \cite{Abt:2004yk,Ackermann:2012xja} is
expected to reach 
\mbox{$T^{0\nu\beta\beta}_{1/2}(^{76}{\rm Ge}) \geq
2\times 10^{26}$ yr}, corresponding to $m_{\beta \beta }\leq 100$ meV. A
bolometric CUORE experiment, using ${}^{130}Te$ \cite{Alessandria:2011rc},
is currently under construction and has an estimated sensitivity close to $%
T_{1/2}^{0\nu \beta \beta }(^{130}\mathrm{Te})\sim 10^{26}$ yr, which
corresponds to \mbox{$m_{\beta\beta}\leq 50$ meV.} Besides that, there are
proposals for ton-scale next-to-next generation $0\nu \beta \beta $
experiments with $^{136}$Xe \cite{KamLANDZen:2012aa,Albert:2014fya} and $%
^{76}$Ge \cite{Abt:2004yk,Guiseppe:2011me} which claim sensitivities over $%
T_{1/2}^{0\nu \beta \beta }\sim 10^{27}$ yr, corresponding to $m_{\beta
\beta }\sim 12-30$ meV. For a recent review, see for example Ref. \cite%
{Bilenky:2014uka}. Consequently, our model predicts $T_{1/2}^{0\nu \beta
\beta }$, which is at the level of the sensitivities of the next generation or
next-to-next generation $0\nu \beta \beta $ experiments.

\section{Quark masses and mixings}

\label{quarkmassesandmixing}

From the quark Yukawa terms of Eq. (\ref{Lyq}), it follows that the SM quark
mass matrices take the form: 
\begin{eqnarray}
M_{U} &=&\frac{v}{\sqrt{2}}\left( 
\begin{array}{ccc}
c_{1}\lambda ^{8} & 0 & a_{1}\lambda ^{4} \\ 
0 & b_{1}\lambda ^{4} & a_{2}\lambda ^{2} \\ 
0 & 0 & a_{3}%
\end{array}%
\right) ,  \label{Quarktextures} \\
M_{D} &=&\frac{v}{\sqrt{2}}\left( 
\begin{array}{ccc}
e_{1}\lambda ^{7} & f_{1}\lambda ^{6} & 0 \\ 
0 & f_{2}\lambda ^{5} & 0 \\ 
0 & 0 & g_{1}\lambda ^{3}%
\end{array}%
\right) ,  \notag
\end{eqnarray}%
where $a_{k}$ ($k=1,2,3$), $b_{1}$, $c_{1}$, $g_{1}$, $f_{1}$, $f_{2}$ and $%
e_{1}$ are $\mathcal{O}(1)$ parameters. Here $\lambda =0.225$ is one of the
Wolfenstein parameters and $v=246$ GeV the scale of electroweak symmetry
breaking. From the SM quark mass textures given above, it follows that the Cabbibo mixing emerges from the down type quark sector, whereas the up type quark sector generates the remaining mixing angles. Besides that, the low energy quark flavor data indicates that the CP violating phase in the quark sector is associated with the quark mixing angle in the 1-3 plane, as follows from the Standard parametrization of the quark mixing matrix. Consequently, in order to get quark mixing angles and a CP violating phase consistent with the experimental data, we assume that all dimensionless parameters given in Eq. (\ref{Quarktextures}) are real, except for $a_{1}$, taken to be complex. 

Furthermore, as follows from the different $\Delta(27)$ singlet assignments for the quark fields, the exotic quarks do not mix with the SM quarks. We find that the exotic quark masses are: 
\begin{eqnarray}
m_{T} &=&y^{\left( T\right) }\frac{v_{\chi }}{\sqrt{2}},\hspace{1cm}
\label{mexotics} \\
m_{J^{1}} &=&y_{1}^{\left( J\right) }\frac{v_{\chi }}{\sqrt{2}}=\frac{%
y_{1}^{\left( J\right) }}{y^{\left( T\right) }}m_{T},\hspace{0.5cm}%
m_{J^{2}}=y_{2}^{\left( J\right) }\frac{v_{\chi }}{\sqrt{2}}=\frac{%
y_{2}^{\left( J\right) }}{y^{\left( T\right) }}m_{T}.  \notag
\end{eqnarray}

Since the the breaking of the $\Delta \left( 27\right) \otimes Z_{4}\otimes
Z_{8}\otimes Z_{14}$ discrete group gives rise to the observed pattern of
charged fermion masses and quark mixing angles, and in order to simplify the
analysis, we set $e_{1}=f_{1}$ as well as$\ c_{1}=a_{3}=1$ and $g_{1}=b_{1}$%
, motivated by naturalness arguments and by the relation $m_{c}\sim m_{b}$,
respectively. Consequently, there are only 6 effective free parameters in
the SM quark sector of our model, i.e., $\left\vert a_{1}\right\vert $, $%
a_{2}$, $b_{1}$, $f_{1}$, $f_{2}$ and the phase $\gamma _{q}$. We fit these
6 parameters to reproduce the 10 physical observables of the quark sector,
i.e., the six quark masses, the three mixing angles and the CP violating
phase. By varying the parameters $\left\vert a_{1}\right\vert $, $a_{2}$, $%
b_{1}$, $f_{1}$, $f_{2}$ and $\gamma _{q}$, we find the quark masses, the
three quark mixing angles and the CP violating phase $\delta $ reported in
Table \ref{Tab}, which correspond to the best fit values: 
\begin{eqnarray}
\left\vert a_{1}\right\vert &\simeq &1.36,\hspace{1cm}a_{2}\simeq 0.80,%
\hspace{1cm}b_{1}\simeq 1.43,  \notag \\
f_{1} &\simeq &0.58,\hspace{1cm}f_{2}\simeq 0.57,\hspace{1cm}\gamma
_{q}=-112^{\circ }.
\end{eqnarray}

\begin{table}[tbh]
\begin{center}
\begin{tabular}{c|l|l}
\hline\hline
Observable & Model value & Experimental value \\ \hline
$m_{u}(MeV)$ & \quad $1.16$ & \quad $1.45_{-0.45}^{+0.56}$ \\ \hline
$m_{c}(MeV)$ & \quad $641$ & \quad $635\pm 86$ \\ \hline
$m_{t}(GeV)$ & \quad $174$ & \quad $172.1\pm 0.6\pm 0.9$ \\ \hline
$m_{d}(MeV)$ & \quad $2.9$ & \quad $2.9_{-0.4}^{+0.5}$ \\ \hline
$m_{s}(MeV)$ & \quad $59.2$ & \quad $57.7_{-15.7}^{+16.8}$ \\ \hline
$m_{b}(GeV)$ & \quad $2.85$ & \quad $2.82_{-0.04}^{+0.09}$ \\ \hline
$\sin \theta _{12}$ & \quad $0.225$ & \quad $0.225$ \\ \hline
$\sin \theta _{23}$ & \quad $0.0407$ & \quad $0.0412$ \\ \hline
$\sin \theta _{13}$ & \quad $0.00352$ & \quad $0.00351$ \\ \hline
$\delta $ & \quad $68^{\circ }$ & \quad $68^{\circ }$ \\ \hline\hline
\end{tabular}%
\end{center}
\caption{Model and experimental values of the quark masses and CKM
parameters.}
\label{Tab}
\end{table}
In Table \ref{Tab} we show the model and experimental values for the
physical observables of the quark sector. We use the $M_{Z}$-scale
experimental values of the quark masses given by Ref. \cite{Bora:2012tx}
(which are similar to those in \cite{Xing:2007fb}). The experimental values
of the CKM parameters are taken from Ref. \cite{Agashe:2014kda}. As
indicated by Table \ref{Tab}, the obtained quark masses, quark mixing angles, and CP violating phase are highly consistent with the experimental low
energy quark flavor data. Note that in our previous paper \cite{Vien:2016tmh}%
, the CKM matrix, at the tree level, is the identity, which should be
improved by higher order loop corrections.

\section{Conclusions}

\label{conclusions}

We constructed the first multiscalar singlet extension of the original 3-3-1 model with right-handed neutrinos, based on the $\Delta \left( 27\right) $ family
symmetry supplemented by the $Z_{4}\otimes Z_{8}\otimes Z_{14}$ discrete
group. Contrary to the previous $\Delta(27)$ flavor 3-3-1 model \cite{Vien:2016tmh}, where
the CKM matrix is the identity, this model provides an excellent description
of the observed SM fermion mass and mixing pattern. The $\Delta \left(
27\right) $, $Z_{4}$, and $Z_{8}$ symmetries allow one to reduce the number of
parameters in the Yukawa terms, increasing the predictivity power of the
model, whereas the $Z_{14}$ symmetry causes the charged fermion mass and
quark mixing pattern. In the model under consideration, the light active
neutrino masses are generated from a double seesaw mechanism and the
observed pattern of charged fermion masses and quark mixing angles is caused
by the breaking of the $\Delta \left( 27\right) \otimes Z_{4}\otimes
Z_{8}\otimes Z_{14}$ discrete group at very high energy. The resulting the
neutrino spectrum of our model is composed of light active neutrinos, heavy
and very heavy sterile neutrinos. The smallness of the active neutrino
masses arises from their scaling with inverse powers of the large model
cutoff $\Lambda $ and by their quadratic dependence on the very small vacuum
expectation value of the $\Delta \left( 27\right) $ scalar triplets $\Omega $
and $\Theta $ participating in the Dirac neutrino Yukawa interactions. The
SM Yukawa sector of our predictive $\Delta \left( 27\right) $ flavor 3-3-1
model has in total only 14 effective free parameters (8 and 6 effective free
parameters in the lepton and quark sectors, respectively), which are fitted
to reproduce the experimental values of the 18 physical observables in the
quark and lepton sectors, i.e., 9 charged fermion masses, 2 neutrino mass
squared splittings, 3 lepton mixing parameters, 3 quark mixing angles and 1
CP violating phase of the CKM quark mixing matrix. The obtained physical
observables for the quark sector are consistent with the experimental data,
whereas the ones for the lepton also do but only for the inverted neutrino
mass hierarchy. The normal neutrino mass hierarchy scenario of our model is
disfavored by the neutrino oscillation experimental data. We find an
effective Majorana neutrino mass parameter of neutrinoless double beta decay
of $m_{\beta \beta }=$ 22 meV, a leptonic Dirac CP violating phase of $%
34^{\circ }$ and a Jarlskog invariant of about $10^{-2}$ for the inverted
neutrino mass spectrum. Our obtained value of 22 meV for the effective
Majorana neutrino mass is within the declared reach of the next generation
bolometric CUORE experiment \cite{Alessandria:2011rc} or, more
realistically, of the next-to-next generation ton-scale $0\nu \beta \beta $%
-decay experiments.

\subsection*{Acknowledgements}

This research has received funding from the Vietnam National Foundation for
Science and Technology Development (NAFOSTED) under Grant number 103.01-2014.51. A.E.C.H was supported by DGIP internal Grant No. 111458. H. N. Long thanks
Universidad T\'{e}cnica Federico Santa Mar\'{\i}a for hospitality, where
this work was finished. The visit of H. N. Long to Universidad T\'{e}cnica
Federico Santa Mar\'{\i}a was supported by DGIP internal Grant No. 111458.

\section*{Appendices}

\appendix


\section{The product rules of the $\Delta (27)$ discrete group}

\label{A}

The $\Delta (27)$ discrete group is a subgroup of $SU(3)$, has 27 elements
divided into 11 conjugacy classes. Then the $\Delta (27)$ discrete group
contains the following 11 irreducible representations: two triplets, i.e., $%
\mathbf{3}_{[0][1]}$ (which we denote by $\mathbf{3}$)\ and its conjugate $%
\mathbf{3}_{[0][2]}$ (which we denote by $\overline{\mathbf{3}}$) and 9
singlets, i.e., $\mathbf{1}_{k,l}$ ($k,l=0,1,2$), where $k$ and $l$
correspond to the $Z_{3}$ and $Z_{3}^{\prime }$ charges, respectively \cite%
{Ishimori:2010au}. The $\Delta (27)$ discrete group, which is a simple group
of the type $\Delta (3n^{2})$ with $n=3$, is isomorphic to the semi-direct
product group $(Z_{3}^{\prime }\times Z_{3}^{\prime \prime })\rtimes Z_{3}$ 
\cite{Ishimori:2010au}. It is worth mentioning that the simplest group of
the type $\Delta (3n^{2})$ is $\Delta (3)\equiv Z_{3}$. The next group is $%
\Delta (12)$, which is isomorphic to $A_{4}$. Consequently the $\Delta (27)$
discrete group is the simplest nontrivial group of the type $\Delta(3n^{2})$%
. Any element of the $\Delta (27)$ discrete group can be expressed as $%
b^{k}a^{m}{a^{\prime }}^{n}$, being $b$, $a$ and $a^{\prime }$ the generators of the $Z_{3}$, $Z_{3}^{\prime }$ and $Z_{3}^{\prime \prime }$
cyclic groups, respectively. These generators fulfill the relations: 
\begin{eqnarray}
&& a^3 =a^{\prime 3}=b^3=1,\hspace*{0.5cm} a a^{\prime }=a^{\prime }a, 
\notag \\
&& bab^{-1}=a^{-1}a^{\prime -1},\,\, b a^{\prime}b^{-1}=a,
\label{aapbrela}
\end{eqnarray}%
The characters of the $\Delta (27)$ discrete group are shown in Table \ref%
{tab:delta-27}. Here $n$ is the number of elements, $h$ is the order of each
element, and $\omega =e^{\frac{2\pi i}{3}}=-\frac{1}{2}+i\frac{\sqrt{3}}{2}$
is the cube root of unity, which satisfies the relations $1+\omega +\omega
^{2}=0$ and $\omega ^{3}=1$.
The conjugacy classes of $\Delta (27)$ are given by: 
\begin{equation*}
\begin{array}{ccc}
C_{1}: & \{e\}, & h=1, \\ 
C_{1}^{(1)}: & \{a,a^{\prime 2}\}, & h=3, \\ 
C_{1}^{(2)}: & \{a^{2},a^{\prime }\}, & h=3, \\ 
C_{3}^{(0,1)}: & \{a^{\prime 2}a^{\prime 2}\}, & h=3, \\ 
C_{3}^{(0,2)}: & \{a^{\prime 2},a^{2},aa^{\prime }\}, & h=3, \\ 
C_{3}^{(1,p)}: & \{ba^{p},ba^{p-1}a^{\prime p-2}a^{\prime 2}\}, & h=3, \\ 
C_{3}^{(2,p)}: & \{ba^{p},ba^{p-1}a^{\prime p-2}a^{\prime 2}\}, & h=3. \\ 
&  & 
\end{array}%
\end{equation*}

\begin{table}[t]
\begin{center}
\begin{tabular}{|c|c|c|c|c|}
\hline
& h & $\chi_{1_{(r,s)}}$ & $\chi_{3_{[0,1]}}$ & $\chi_{3_{[0,2]}}$ \\ \hline
$1C_1$ & 1 & 1 & 3 & 3 \\ \hline
$1C_1^{(1)}$ & 1 & 1 & $3\omega^2$ & $3\omega$ \\ \hline
$1C_1^{(2)}$ & 1 & 1 & $3\omega$ & $3\omega^2$ \\ \hline
$3C_1^{(0,1)}$ & $3$ & $\omega^{s}$ & $0$ & $0$ \\ \hline
$3C_1^{(0,2)}$ & $3$ & $\omega^{2s}$ & $0$ & $0$ \\ \hline
$C_3^{(1,p)}$ & $3$ & $\omega^{r+s p}$ & 0 & 0 \\ \hline
$C_3^{(2,p)}$ & $3$ & $\omega^{2r+s p}$ & 0 & 0 \\ \hline
\end{tabular}%
\end{center}
\caption{Characters of $\Delta (27)$}
\label{tab:delta-27}
\end{table}
The tensor products between $\Delta (27)$ triplets are described by the
following relations \cite{Ishimori:2010au}: 
\begin{eqnarray}\vspace{-1cm}
\begin{pmatrix}
x_{1,-1} \\ 
x_{0,1} \\ 
x_{-1,0} \\ 
\end{pmatrix}%
_{\mathbf{3}_{[0][1]}}\otimes 
\begin{pmatrix}
y_{1,-1} \\ 
y_{0,1} \\ 
y_{-1,0} \\ 
\end{pmatrix}%
_{\mathbf{3}_{[0][1]}} &=&%
\begin{pmatrix}
x_{1,-1}y_{1,-1} \\ 
x_{0,1}y_{0,1} \\ 
x_{-1,0}y_{-1,0} \\ 
\end{pmatrix}%
_{\mathbf{3}_{[0][2]}^{\left( S_{1}\right) }}\oplus \frac{1}{2}%
\begin{pmatrix}
x_{0,1}y_{-1,0}+x_{-1,0}y_{0,1} \\ 
x_{-1,0}y_{1,-1}+x_{1,-1}y_{-1,0} \\ 
x_{1,-1}y_{0,1}+x_{0,1}y_{1,-1} \\ 
\end{pmatrix}%
_{\mathbf{3}_{[0][2]}^{\left( S_{2}\right) }}  \notag \\
&&\oplus \frac{1}{2}%
\begin{pmatrix}
x_{0,1}y_{-1,0}-x_{-1,0}y_{0,1} \\ 
x_{-1,0}y_{1,-1}-x_{1,-1}y_{-1,0} \\ 
x_{1,-1}y_{0,1}-x_{0,1}y_{1,-1} \\ 
\end{pmatrix}%
_{\mathbf{3}_{[0][2]}^{\left( A\right) }},\\
\begin{pmatrix}
x_{2,-2} \\ 
x_{0,2} \\ 
x_{-2,0} \\ 
\end{pmatrix}%
_{\mathbf{3}_{[0][2]}}\otimes 
\begin{pmatrix}
y_{2,-2} \\ 
y_{0,2} \\ 
y_{-2,0} \\ 
\end{pmatrix}%
_{\mathbf{3}_{[0][2]}} &=&%
\begin{pmatrix}
x_{2,-2}y_{2,-2} \\ 
x_{0,2}y_{0,2} \\ 
x_{-2,0}y_{-2,0} \\ 
\end{pmatrix}%
_{\mathbf{3}_{[0][1]}^{\left( S_{1}\right) }}\oplus \frac{1}{2}%
\begin{pmatrix}
x_{0,2}y_{-2,0}+x_{-2,0}y_{0,2} \\ 
x_{-2,0}y_{2,-2}+x_{2,-2}y_{-2,0} \\ 
x_{2,-2}y_{0,2}+x_{0,2}y_{2,-2} \\ 
\end{pmatrix}%
_{\mathbf{3}_{[0][1]}^{\left( S_{2}\right) }}  \notag \\
&&
\oplus \frac{1}{2}%
\begin{pmatrix}
x_{0,2}y_{-2,0}-x_{-2,0}y_{0,2} \\ 
x_{-2,0}y_{2,-2}-x_{2,-2}y_{-2,0} \\ 
x_{2,-2}y_{0,2}-x_{0,2}y_{2,-2} \\ 
\end{pmatrix}%
_{\mathbf{3}_{[0][1]}^{\left( A\right) }},\\
\begin{pmatrix}
x_{1,-1} \\ 
x_{0,1} \\ 
x_{-1,0} \\ 
\end{pmatrix}%
_{\mathbf{3}_{[0][1] }} \otimes 
\begin{pmatrix}
y_{ -1,1} \\ 
y_{0, -1} \\ 
y_{1, 0} \\ 
\end{pmatrix}%
_{\mathbf{3}_{[0][2] }} =&& \sum_r(x_{1,-1}y_{ -1,1}
+\omega^{2r}x_{0,1}y_{0, -1} +\omega^{r}x_{-1,0}y_{1, 0})_{\mathbf{1}%
_{(r,0)}}  \notag \\
&\oplus& \sum_r(x_{1,-1}y_{ 0,-1} +\omega^{2r}x_{0,1}y_{1,0}
+\omega^{r}x_{-1,0}y_{-1,1})_{\mathbf{1}_{(r,1)}}  \notag \\
&\oplus& \sum_r(x_{1,-1}y_{ 1,0} +\omega^{2r}x_{0,1}y_{-1,1}
+\omega^{r}x_{-1,0}y_{0, -1})_{\mathbf{1}_{(r,2)}}.\notag  \\
\end{eqnarray}
The multiplication rules between $\Delta (27)$ singlets and $\Delta (27)$
triplets are given by \cite{Ishimori:2010au}:
\begin{eqnarray}
&&%
\begin{pmatrix}
x_{(1,-1)} \\ 
x_{(0,1)} \\ 
x_{(-1,0)}%
\end{pmatrix}%
_{\mathbf{3}_{[0][1]}}\otimes (z)_{1_{k,l}}=%
\begin{pmatrix}
x_{(1,-1)}z \\ 
\omega ^{r}x_{(0,1)}z \\ 
\omega ^{2r}x_{(-1,0)}z%
\end{pmatrix}%
_{\mathbf{3}_{[l][1+l]}},   \\
&&%
\begin{pmatrix}
x_{(2,-2)} \\ 
x_{(0,2)} \\ 
x_{(-2,0)}%
\end{pmatrix}%
_{\mathbf{3}_{[0][2]}}\otimes (z)_{1_{k,l}}=%
\begin{pmatrix}
x_{(2,-2)}z \\ 
\omega ^{r}x_{(0,2)}z \\ 
\omega ^{2r}x_{(-2,0)}%
\end{pmatrix}%
_{\mathbf{3}_{[l][2+l]}}.
\end{eqnarray}%
The tensor products of $\Delta (27)$ singlets $\mathbf{1}_{k,\ell }$ and $%
\mathbf{1}_{k^{\prime },\ell ^{\prime }}$ take the form \cite%
{Ishimori:2010au}: 
\begin{equation}
\mathbf{1}_{k,\ell }\otimes \mathbf{1}_{k^{\prime },\ell ^{\prime }}=\mathbf{%
1}_{k+k^{\prime }\func{mod}3,\ell +\ell ^{\prime }\func{mod}3}.
\end{equation}
From the equation given above, we obtain explicitly the singlet
multiplication rules of the $\Delta(27)$ group, which are given in Table \ref{D27multiplets}.
\begin{table}[tbh]
\begin{center}
\begin{tabular}{|c||c|c|c|c|c|c|c|c|}
\hline
Singlets & ~ $\mathbf{1}_{01}$ ~ & ~ $\mathbf{1}_{02}$ ~ & ~ $\mathbf{1}%
_{10} $ ~ & ~ $\mathbf{1}_{11}$~ & ~ $\mathbf{1}_{12}$ ~ & ~ $\mathbf{1}%
_{20} $ ~ & ~ $\mathbf{1}_{21}$~ & $\mathbf{1}_{22}$  \\ \hline\hline
$\mathbf{1}_{01}$ & $\mathbf{1}_{02}$ & $\mathbf{1}_{00}$ & $\mathbf{1}_{11}$
& $\mathbf{1}_{12}$ & $\mathbf{1}_{10}$ & $\mathbf{1}_{21}$ & $\mathbf{1}%
_{22}$ & $\mathbf{1}_{20}$  \\ \hline
$\mathbf{1}_{02}$ & $\mathbf{1}_{00}$ & $\mathbf{1}_{01}$ & $\mathbf{1}_{12}$
& $\mathbf{1}_{10}$ & $\mathbf{1}_{11}$ & $\mathbf{1}_{22}$ & $\mathbf{1}%
_{20}$ & $\mathbf{1}_{21}$ \\ \hline
$\mathbf{1}_{10}$ & $\mathbf{1}_{11}$ & $\mathbf{1}_{12}$ & $\mathbf{1}_{20}$
& $\mathbf{1}_{21}$ & $\mathbf{1}_{22}$ & $\mathbf{1}_{00}$ & $\mathbf{1}%
_{01}$ & $\mathbf{1}_{02}$  \\ \hline
$\mathbf{1}_{11}$ & $\mathbf{1}_{12}$ & $\mathbf{1}_{10}$ & $\mathbf{1}_{21}$
& $\mathbf{1}_{22}$ & $\mathbf{1}_{20}$ & $\mathbf{1}_{01}$ & $\mathbf{1}%
_{02}$ & $\mathbf{1}_{00}$  \\ \hline
$\mathbf{1}_{12}$ & $\mathbf{1}_{10}$ & $\mathbf{1}_{11}$ & $\mathbf{1}_{22}$
& $\mathbf{1}_{20}$ & $\mathbf{1}_{21}$ & $\mathbf{1}_{02}$ & $\mathbf{1}%
_{00}$ & $\mathbf{1}_{01}$  \\ \hline
$\mathbf{1}_{20}$ & $\mathbf{1}_{21}$ & $\mathbf{1}_{22}$ & $\mathbf{1}_{00}$
& $\mathbf{1}_{01}$ & $\mathbf{1}_{02}$ & $\mathbf{1}_{10}$ & $\mathbf{1}%
_{11}$ & $\mathbf{1}_{12}$  \\ \hline
$\mathbf{1}_{21}$ & $\mathbf{1}_{22}$ & $\mathbf{1}_{20}$ & $\mathbf{1}_{01}$
& $\mathbf{1}_{02}$ & $\mathbf{1}_{00}$ & $\mathbf{1}_{11}$ & $\mathbf{1}%
_{12}$ & $\mathbf{1}_{10}$ \\ \hline
$\mathbf{1}_{22}$ & $\mathbf{1}_{20}$ & $\mathbf{1}_{21}$ & $\mathbf{1}_{02}$
& $\mathbf{1}_{00}$ & $\mathbf{1}_{01}$ & $\mathbf{1}_{12}$ & $\mathbf{1}%
_{10}$ & $\mathbf{1}_{11}$  \\ \hline
\end{tabular}%
\end{center}
\caption{The singlet multiplications of the group $\Delta (27)$.}
\label{D27multiplets}
\end{table}

\section{Scalar potential for two $\Delta (27)$ scalar triplets}
\label{ApB}
The scalar potential for two $\Delta (27)$ scalar triplets, i.e., $U$ and $W$ having different $Z_N$ charges can be written as follows:
\begin{equation}
V=V_{U}+V_{W}+V_{U,W}
\label{scalarpotentialtwotriplets}
\end{equation}
where $V_U$ and $V_w$ are the scalar potentials for the $\Delta (27)$ scalar triplets $U$ and $W$, respectively, whereas $V_{U,W}$ include the interaction terms involving both $\Delta (27)$ scalar triplets $U$ and $W$. The different parts of the scalar potential for the two $\Delta (27)$ scalar triplets are given by:
\begin{eqnarray}
V_{U} &=&-\mu _{U}^{2}\left( UU^{\ast }\right) _{\mathbf{\mathbf{1}_{0%
\mathbf{,0}}}}+\kappa _{U,1}\left( UU^{\ast }\right) _{\mathbf{\mathbf{1}_{0%
\mathbf{,0}}}}\left( UU^{\ast }\right) _{\mathbf{\mathbf{1}_{0\mathbf{,0}}}}
\notag\\
&&+\kappa _{U,2}\left( UU^{\ast }\right) _{\mathbf{\mathbf{1}_{1\mathbf{,0}}}%
}\left( UU^{\ast }\right) _{\mathbf{\mathbf{1}_{2\mathbf{,0}}}}\notag \\
&&+\kappa _{U,3}\left( UU^{\ast }\right) _{\mathbf{\mathbf{1}_{0,1}}}\left(
UU^{\ast }\right) _{\mathbf{\mathbf{1}_{0,2}}}\notag \\
&&+\kappa _{U,4}\left[ \left( UU^{\ast }\right) _{\mathbf{\mathbf{1}_{1,1}}%
}\left( UU^{\ast }\right) _{\mathbf{\mathbf{1}_{2,2}}}+H.c\right]\notag \\
&&+\kappa _{U,5}\left( UU\right) _{\overline{\mathbf{3}}\mathbf{_{S_{1}}}%
}\left( U^{\ast }U^{\ast }\right) _{\mathbf{3}_{S_{1}}}\notag \\
&&+\kappa _{U,6}\left( UU\right) _{\overline{\mathbf{3}}\mathbf{_{S_{2}}}%
}\left( U^{\ast }U^{\ast }\right) _{\mathbf{3}_{S_{2}}} \notag\\
&&+\kappa _{U,7}\left[ \left( UU\right) _{\overline{\mathbf{3}}\mathbf{%
_{S_{1}}}}\left( U^{\ast }U^{\ast }\right) _{\mathbf{3}_{S_{2}}}+H.c\right]
\end{eqnarray}
\begin{equation}
V_{W}=V_{U}\left( U\rightarrow W,\mu _{U}\rightarrow \mu _{W},\kappa
_{U,j}\rightarrow \kappa _{W,j}\right) .
\end{equation}
\begin{eqnarray}
V_{U,W} &=&\gamma _{UW,1}\left( UU^{\ast }\right) _{\mathbf{\mathbf{1}_{0%
\mathbf{,0}}}}\left( WW^{\ast }\right) _{\mathbf{\mathbf{1}_{0\mathbf{,0}}}}
\\
&&+\kappa _{UW,1}\left( UW^{\ast }\right) _{\mathbf{\mathbf{1}_{0\mathbf{,0}}%
}}\left( U^{\ast }W\right) _{\mathbf{\mathbf{1}_{0\mathbf{,0}}}}\notag \\
&&+\gamma _{UW,2}\left[ \left( UU^{\ast }\right) _{\mathbf{\mathbf{1}_{1%
\mathbf{,0}}}}\left( WW^{\ast }\right) _{\mathbf{\mathbf{1}_{2\mathbf{,0}}}%
}+H.c\right]\notag \\
&&+\kappa _{UW,2}\left[ \left( UW^{\ast }\right) _{\mathbf{\mathbf{1}_{1%
\mathbf{,0}}}}\left( UW^{\ast }\right) _{\mathbf{\mathbf{1}_{2\mathbf{,0}}}%
}+H.c\right]\notag \\
&&+\gamma _{UW,3}\left[ \left( UU^{\ast }\right) _{\mathbf{\mathbf{1}_{0,1}}%
}\left( WW^{\ast }\right) _{\mathbf{\mathbf{1}_{0,2}}}+H.c\right]\notag \\
&&+\kappa _{UW,3}\left[ \left( UW^{\ast }\right) _{\mathbf{\mathbf{1}_{0,1}}%
}\left( UW^{\ast }\right) _{\mathbf{\mathbf{1}_{0,2}}}+H.c\right]\notag \\
&&+\gamma _{UW,4}\left[ \left( UU^{\ast }\right) _{\mathbf{\mathbf{1}_{1,1}}%
}\left( WW^{\ast }\right) _{\mathbf{\mathbf{1}_{2,2}}}+H.c\right]\notag \\
&&+\kappa _{UW,4}\left[ \left( UW^{\ast }\right) _{\mathbf{\mathbf{1}_{1,1}}%
}\left( UW^{\ast }\right) _{\mathbf{\mathbf{1}_{2,2}}}+H.c\right]\notag \\
&&+\gamma _{UW,5}\left[ \left( UU\right) _{\overline{\mathbf{3}}\mathbf{%
_{S_{1}}}}\left( W^{\ast }W^{\ast }\right) _{\mathbf{3}_{S_{1}}}+H.c\right]
\notag\\
&&+\gamma _{UW,6}\left[ \left( UU\right) _{\overline{\mathbf{3}}\mathbf{%
_{S_{2}}}}\left( W^{\ast }W^{\ast }\right) _{\mathbf{3}_{S_{2}}}+H.c\right]
\notag\\
&&+\kappa _{UW,5}\left( UW\right) _{\overline{\mathbf{3}}\mathbf{_{S_{1}}}%
}\left( U^{\ast }W^{\ast }\right) _{\mathbf{3}_{S_{1}}}\notag \\
&&+\kappa _{UW,6}\left( UW\right) _{\overline{\mathbf{3}}\mathbf{_{S_{2}}}%
}\left( U^{\ast }W^{\ast }\right) _{\mathbf{3}_{S_{2}}}\notag \\
&&+\gamma _{UW,7}\left[ \left( UU\right) _{\overline{\mathbf{3}}\mathbf{%
_{S_{1}}}}\left( W^{\ast }W^{\ast }\right) _{\mathbf{3}_{S_{2}}}+H.c\right]\notag
\\
&&+\kappa _{UW,7}\left[ \left( UW\right) _{\overline{\mathbf{3}}\mathbf{%
_{S_{1}}}}\left( U^{\ast }W^{\ast }\right) _{\mathbf{3}_{S_{2}}}+H.c\right]
\notag\\
&&+\kappa _{UW,8}\left[ \left( UW\right) _{\overline{\mathbf{3}}\mathbf{_{A}}%
}\left( U^{\ast }W^{\ast }\right) _{\mathbf{3}_{A}}+H.c\right]\notag \\
&&+\kappa _{UW,9}\left[ \left( UW\right) _{\overline{\mathbf{3}}\mathbf{_{A}}%
}\left( U^{\ast }W^{\ast }\right) _{\mathbf{3}_{S_{1}}}+H.c\right]\notag \\
&&+\kappa _{UW,10}\left[ \left( UW\right) _{\overline{\mathbf{3}}\mathbf{_{A}%
}}\left( U^{\ast }W^{\ast }\right) _{\mathbf{3}_{S_{2}}}+H.c\right]\notag
\end{eqnarray}
where the $\Delta(27)$ scalar triplets $U$ and $W$ acquire the following VEV pattern:
\begin{equation}
\left\langle U\right\rangle =\left( u_{1},u_{2},u_{3}\right) ,\hspace*{0.5cm}%
\left\langle W\right\rangle =\left( w_{1},w_{2},w_{3}\right) ,
\end{equation}
Then, from the previous expressions, the following relations are obtained:
\begin{eqnarray}
\frac{\partial V_{U}}{\partial u\text{$_{_{1}}$}} &=&-2u_{1}\mu
_{U}^{2}+4\kappa _{U,1}u_{1}\left( u\text{$_{1}^{2}+$}u\text{$_{2}^{2}+$}u%
\text{$_{3}^{2}$}\right)\notag \\
&&+2\kappa _{U,2}u_{1}\left( 2u_{1}^{2}-u_{2}^{2}-u_{3}^{2}\right)\notag \\
&&+2\kappa _{U,3}\left( u_{2}+u_{3}\right) \left[ u_{1}\left(
u_{2}+u_{3}\right) +u_{2}u_{3}\right]\notag \\
&&+\kappa _{U,4}\left[ 2u_{1}u_{2}u_{3}+u_{2}u_{3}\left( u_{2}+u_{3}\right)
-2u_{1}\left( u_{2}^{2}+u_{3}^{2}\right) \right]\notag \\
&&+2\kappa _{U,5}u_{1}^{3}+2\kappa _{U,6}u_{1}\left(
u_{2}^{2}+u_{3}^{2}\right)\notag \\
&&+2\kappa _{U,7}u_{2}u_{3}\left( 2u_{1}+u_{2}+u_{3}\right) ,\label{dV1}\\
&&\notag \\
\frac{\partial V_{U}}{\partial u\text{$_{2}$}} &=&-2u_{2}\mu
_{U}^{2}+4\kappa _{U,2}u_{2}\left( u\text{$_{1}^{2}+$}u\text{$_{2}^{2}+$}u%
\text{$_{3}^{2}$}\right)\notag \\
&&+2\kappa _{U,2}u_{2}\left( 2u_{2}^{2}-u_{1}^{2}-u_{3}^{2}\right)\notag \\
&&+2\kappa _{U,3}\left( u_{1}+u_{3}\right) \left[ u_{2}\left(
u_{1}+u_{3}\right) +u_{1}u_{3}\right]\notag \\
&&+\kappa _{U,4}\left[ 2u_{1}u_{2}u_{3}+u_{1}u_{3}\left( u_{1}+u_{3}\right)
-2u_{2}\left( u_{1}^{2}+u_{3}^{2}\right) \right]\notag \\
&&+2\kappa _{U,5}u_{2}^{3}+2\kappa _{U,6}u_{2}\left(
u_{1}^{2}+u_{3}^{2}\right)\notag \\
&&+2\kappa _{U,7}u_{1}u_{3}\left( 2u_{2}+u_{1}+u_{3}\right),\label{dV2} \\
&& \notag\\
\frac{\partial V_{U}}{\partial u\text{$_{3}$}} &=&-2u_{3}\mu
_{U}^{2}+4\kappa _{U,2}u_{3}\left( u\text{$_{1}^{2}+$}u\text{$_{2}^{2}+$}u%
\text{$_{3}^{2}$}\right)\notag \\
&&+2\kappa _{U,2}u_{3}\left( 2u_{3}^{2}-u_{2}^{2}-u_{3}^{2}\right)\notag \\
&&+2\kappa _{U,3}\left( u_{1}+u_{2}\right) \left[ u_{3}\left(
u_{1}+u_{2}\right) +u_{1}u_{2}\right]\notag \\
&&+\kappa _{U,4}\left[ 2u_{1}u_{2}u_{3}+u_{1}u_{2}\left( u_{1}+u_{2}\right)
-2u_{3}\left( u_{1}^{2}+u_{2}^{2}\right) \right]\notag \\
&&+2\kappa _{U,5}u_{3}^{3}+2\kappa _{U,6}u_{3}\left(
u_{1}^{2}+u_{2}^{2}\right)\notag \\
&&+2\kappa _{U,7}u_{1}u_{2}\left( 2u_{3}+u_{1}+u_{2}\right)\label{dV3} ,
\end{eqnarray}
\begin{eqnarray}
\frac{\partial V_{W}}{\partial w\text{$_{n}$}} &=&\lim_{\mu _{U}\rightarrow
\mu _{W},\kappa _{U,j}\rightarrow \kappa _{W,j},u_{n}\rightarrow w_{n}}\frac{%
\partial V_{U}}{\partial u_{n}},\hspace*{0.5cm}\notag \\
n &=&1,2,3,\hspace*{0.5cm}j=1,\ldots ,10.
\end{eqnarray}
\begin{eqnarray}
\frac{\partial V_{U,W}}{\partial u\text{$_{_{1}}$}} &=&2\gamma
_{UW,1}u_{1}\left( w\text{$_{1}^{2}+w_{2}^{2}+w_{3}^{2}$}\right)\notag \\
&&+2\kappa _{UW,1}w_{1}\left( u_{1}w_{1}+u_{2}w_{2}+u_{3}w_{3}\right)\notag \\
&&+2\gamma _{UW,2}u_{1}\left( 2w_{1}^{2}-w_{2}^{2}-\text{$w_{3}^{2}$}\right)\notag
\\
&&+2\kappa _{UW,2}w_{1}\left( 2u_{1}w_{1}-u_{2}w_{2}-u_{3}w_{3}\right)\notag \\
&&+2\gamma _{UW,3}\left( u_{2}+u_{3}\right) \left[ w_{2}w_{3}+w_{1}\left(
w_{2}+w_{3}\right) \right]\notag \\
&&+2\kappa _{UW,3}\left[ 2u_{1}w_{2}w_{3}+u_{2}\left(
w_{1}w_{2}+w_{3}^{2}\right) \right.\notag \\
&&+\left. u_{3}\left( w_{1}w_{3}+w_{2}^{2}\right) \right]\notag \\
&&+\gamma _{UW,4}\left\{ u_{2}\left[ 2w_{1}w_{3}-w_{2}\left(
w_{1}+w_{3}\right) \right] \right.\notag \\
&&+\left. u_{3}\left[ 2w_{2}w_{3}-w_{1}\left( w_{2}+w_{3}\right) \right]
\right\}\notag \\
&&+\kappa _{UW,4}\left\{ 4u_{1}w_{2}w_{3}-u_{2}\left(
w_{1}w_{2}+w_{3}^{2}\right) \right.\notag \\
&&-\left. u_{3}\left( w_{1}w_{3}+w_{2}^{2}\right) \right\}\notag \\
&&+4\gamma _{UW,5}u_{1}w_{1}^{2}+2\gamma _{UW,6}w_{1}\left(
u_{2}w_{2}+u_{3}w_{3}\right)\notag \\
&&+2\kappa _{UW,5}u_{1}w_{1}^{2}+4\gamma _{UW,7}u_{1}w_{2}w_{3}\notag \\
&&+\frac{\kappa _{UW,6}}{2}\left[ u_{1}\left( w_{2}^{2}+w_{3}^{2}\right)
+w_{1}\left( u_{2}w_{2}+u_{3}w_{3}\right) \right]\notag \\
&&+\kappa _{UW,7}\left[ u_{2}w_{3}\left( w_{1}+w_{2}\right)
+u_{3}w_{2}\left( w_{1}+w_{3}\right) \right]\notag \\
&&+\kappa _{UW,8}\left[ u_{1}\left( w_{2}^{2}+w_{3}^{2}\right) -w_{1}\left(
u_{2}w_{2}+u_{3}w_{3}\right) \right]\notag \\
&&+\kappa _{UW,9}\left[ u_{2}w_{3}\left( w_{1}-w_{2}\right)
+u_{3}w_{2}\left( w_{3}-w_{1}\right) \right]\notag \\
&&+\kappa _{UW,10}u_{1}\left( w_{2}^{2}-w_{3}^{2}\right) ,
\end{eqnarray}
\begin{eqnarray}
\frac{\partial V_{U,W}}{\partial u\text{$_{_{2}}$}} &=&2\gamma
_{UW,1}u_{2}\left( w\text{$_{1}^{2}+w_{2}^{2}+w_{3}^{2}$}\right)\notag \\
&&+2\kappa _{UW,1}w_{2}\left( u_{1}w_{1}+u_{2}w_{2}+u_{3}w_{3}\right)\notag \\
&&+2\gamma _{UW,2}u_{2}\left( 2w_{2}^{2}-w_{1}^{2}-\text{$w_{3}^{2}$}\right)\notag
\\
&&+2\kappa _{UW,2}w_{2}\left( 2u_{2}w_{2}-u_{1}w_{1}-u_{3}w_{3}\right)\notag \\
&&+2\gamma _{UW,3}\left( u_{1}+u_{3}\right) \left[ w_{2}w_{3}+w_{1}\left(
w_{2}+w_{3}\right) \right]\notag \\
&&+2\kappa _{UW,3}\left[ 2u_{2}w_{1}w_{3}+u_{1}\left(
w_{1}w_{2}+w_{3}^{2}\right) \right.\notag \\
&&+\left. u_{3}\left( w_{2}w_{3}+w_{1}^{2}\right) \right]\notag \\
&&+\gamma _{UW,4}\left\{ u_{3}\left[ 2w_{1}w_{2}-w_{3}\left(
w_{1}+w_{2}\right) \right] \right.\notag \\
&&+\left. u_{1}\left[ 2w_{1}w_{3}-w_{2}\left( w_{1}+w_{3}\right) \right]
\right\}\notag \\
&&+\kappa _{UW,4}\left\{ 4u_{2}w_{1}w_{3}-u_{1}\left(
w_{1}w_{2}+w_{3}^{2}\right) \right.\notag \\
&&-\left. u_{3}\left( w_{2}w_{3}+w_{1}^{2}\right) \right\}\notag \\
&&+4\gamma _{UW,5}u_{2}w_{2}^{2}+2\gamma _{UW,6}w_{2}\left(
u_{1}w_{1}+u_{3}w_{3}\right)\notag \\
&&+2\kappa _{UW,5}u_{2}w_{2}^{2}+4\gamma _{UW,7}u_{2}w_{1}w_{3}\notag \\
&&+\frac{\kappa _{UW,6}}{2}\left[ u_{2}\left( w_{1}^{2}+w_{3}^{2}\right)
+w_{2}\left( u_{1}w_{1}+u_{3}w_{3}\right) \right]\notag \\
&&+\kappa _{UW,7}\left[ u_{1}w_{3}\left( w_{1}+w_{2}\right)
+u_{3}w_{1}\left( w_{2}+w_{3}\right) \right]\notag \\
&&+\kappa _{UW,8}\left[ u_{2}\left( w_{1}^{2}+w_{3}^{2}\right) -w_{2}\left(
u_{1}w_{1}+u_{3}w_{3}\right) \right]\notag \\
&&+\kappa _{UW,9}\left[ u_{1}w_{3}\left( w_{1}-w_{2}\right)
+u_{3}w_{1}\left( w_{2}-w_{3}\right) \right]\notag \\
&&+\kappa _{UW,10}u_{2}\left( w_{3}^{2}-w_{1}^{2}\right) ,
\end{eqnarray}
\begin{eqnarray}
\frac{\partial V_{U,W}}{\partial u\text{$_{_{3}}$}} &=&2\gamma
_{UW,1}u_{3}\left( w\text{$_{1}^{2}+w_{2}^{2}+w_{3}^{2}$}\right)\notag \\
&&+2\kappa _{UW,1}w_{3}\left( u_{1}w_{1}+u_{2}w_{2}+u_{3}w_{3}\right)\notag \\
&&+2\gamma _{UW,2}u_{3}\left( 2w_{3}^{2}-w_{1}^{2}-\text{$w_{2}^{2}$}\right)\notag
\\
&&+2\kappa _{UW,2}w_{3}\left( 2u_{3}w_{3}-u_{1}w_{1}-u_{2}w_{2}\right)\notag \\
&&+2\gamma _{UW,3}\left( u_{1}+u_{2}\right) \left[ w_{2}w_{3}+w_{1}\left(
w_{2}+w_{3}\right) \right]\notag \\
&&+2\kappa _{UW,3}\left[ 2u_{3}w_{1}w_{2}+u_{1}\left(
w_{1}w_{3}+w_{2}^{2}\right) \right. \notag\\
&&+\left. u_{2}\left( w_{2}w_{3}+w_{1}^{2}\right) \right]\notag \\
&&+\gamma _{UW,4}\left\{ u_{2}\left[ 2w_{1}w_{2}-w_{3}\left(
w_{1}+w_{2}\right) \right] \right.\notag \\
&&+\left. u_{1}\left[ 2w_{2}w_{3}-w_{1}\left( w_{2}+w_{3}\right) \right]
\right\} \notag\\
&&+\kappa _{UW,4}\left\{ 4u_{3}w_{1}w_{2}-u_{1}\left(
w_{1}w_{3}+w_{2}^{2}\right) \right. \notag\\
&&-\left. u_{2}\left( w_{2}w_{3}+w_{1}^{2}\right) \right\}\notag \\
&&+4\gamma _{UW,5}u_{3}w_{3}^{2}+2\gamma _{UW,6}w_{3}\left(
u_{1}w_{1}+u_{2}w_{2}\right)\notag\\
&&+2\kappa _{UW,5}u_{3}w_{3}^{2}+2\gamma _{UW,7}u_{3}w_{1}w_{2}\notag \\
&&+\frac{\kappa _{UW,6}}{2}\left[ w_{1}\left( u_{2}^{2}+u_{3}^{2}\right)
+u_{1}\left( u_{2}w_{2}+u_{3}w_{3}\right) \right]\notag \\
&&+\kappa _{UW,7}\left[ u_{1}w_{2}\left( w_{1}+w_{3}\right)
+u_{2}w_{1}\left( w_{2}+w_{3}\right) \right]\notag \\
&&+\kappa _{UW,8}\left[ u_{3}\left( w_{1}^{2}+w_{2}^{2}\right) -w_{3}\left(
u_{1}w_{1}+u_{2}w_{2}\right) \right]\notag \\
&&+\kappa _{UW,9}\left[ u_{1}w_{2}\left( w_{3}-w_{1}\right)
+u_{2}w_{1}\left( w_{2}-w_{3}\right) \right] \notag\\
&&+\kappa _{UW,10}u_{3}\left( w_{1}^{2}-w_{2}^{2}\right) ,
\end{eqnarray}%
\begin{eqnarray}
\frac{\partial V_{U,W}}{\partial w\text{$_{_{1}}$}} &=&2\gamma
_{UW,1}w_{1}\left( u\text{$_{1}^{2}+u_{2}^{2}+u_{3}^{2}$}\right)\notag \\
&&+2\kappa _{UW,1}u_{1}\left( u_{1}w_{1}+u_{2}w_{2}+u_{3}w_{3}\right)\notag \\
&&+2\gamma _{UW,2}w_{1}\left( 2u_{1}^{2}-u_{2}^{2}-u\text{$_{3}^{2}$}\right)\notag\\
&&+2\kappa _{UW,2}u_{1}\left( 2u_{1}w_{1}-u_{2}w_{2}-u_{3}w_{3}\right)\notag \\
&&+2\gamma _{UW,3}\left( w_{2}+w_{3}\right) \left[ u_{2}u_{3}+u_{1}\left(
u_{2}+u_{3}\right) \right]\notag \\
&&+2\kappa _{UW,3}\left[ 2w_{1}u_{2}u_{3}+w_{2}\left(
u_{1}u_{2}+u_{3}^{2}\right) \right.\notag \\
&&+\left. w_{3}\left( u_{1}u_{3}+u_{2}^{2}\right) \right]\notag \\
&&+\gamma _{UW,4}\left\{ u_{2}u_{3}\left( 2w_{2}-w_{3}\right)
+u_{1}u_{2}\left( 2w_{3}-w_{2}\right) \right.\notag \\
&&-\left. u_{1}u_{3}\left( w_{2}+w_{3}\right) \right\}\notag \\
&&+\kappa _{UW,4}\left\{ u_{3}\left( 2u_{3}w_{2}-u_{1}w_{3}\right)
+2u_{2}^{2}w_{3}\right.\notag \\
&&-\left. u_{2}\left( 2u_{3}w_{1}+u_{1}w_{2}\right) \right\}\notag \\
&&+4\gamma _{UW,5}w_{1}u_{1}^{2}+2\gamma _{UW,6}u_{1}\left(
u_{2}w_{2}+u_{3}w_{3}\right)\notag \\
&&+2\kappa _{UW,5}w_{1}u_{1}^{2}+2\gamma _{UW,7}\left(
u_{3}^{2}w_{2}+u_{2}^{2}w_{3}\right)\notag \\
&&+\frac{\kappa _{UW,6}}{2}\left[ w_{1}\left( u_{2}^{2}+u_{3}^{2}\right)
+u_{1}\left( u_{2}w_{2}+u_{3}w_{3}\right) \right]\notag \\
&&+\kappa _{UW,7}\left[ w_{2}u_{3}\left( u_{1}+u_{2}\right)
+w_{3}u_{2}\left( u_{1}+u_{3}\right) \right]\notag \\
&&+\kappa _{UW,8}\left[ w_{1}\left( u_{2}^{2}+u_{3}^{2}\right) -u_{1}\left(
u_{2}w_{2}+u_{3}w_{3}\right) \right]\notag \\
&&+\kappa _{UW,9}\left[ w_{2}u_{3}\left( u_{2}-u_{1}\right)
+w_{3}u_{2}\left( u_{1}-u_{3}\right) \right]\notag \\
&&+\kappa _{UW,10}w_{1}\left( u_{3}^{2}-u_{2}^{2}\right) ,
\end{eqnarray}
\begin{eqnarray}
\frac{\partial V_{U,W}}{\partial w\text{$_{_{2}}$}} &=&2\gamma
_{UW,1}w_{2}\left( u\text{$_{1}^{2}+u_{2}^{2}+u_{3}^{2}$}\right)\notag \\
&&+2\kappa _{UW,1}u_{2}\left( u_{1}w_{1}+u_{2}w_{2}+u_{3}w_{3}\right)\notag \\
&&+2\gamma _{UW,2}w_{2}\left( 2u_{2}^{2}-u_{1}^{2}-u\text{$_{3}^{2}$}\right)\notag
\\
&&+2\kappa _{UW,2}u_{2}\left( 2u_{2}w_{2}-u_{1}w_{1}-u_{3}w_{3}\right)\notag \\
&&+2\gamma _{UW,3}\left( w_{1}+w_{3}\right) \left[ u_{2}u_{3}+u_{1}\left(
u_{2}+u_{3}\right) \right]\notag \\
&&+2\kappa _{UW,3}\left[ 2w_{2}u_{1}u_{3}+w_{1}\left(
u_{1}u_{2}+u_{3}^{2}\right) \right.\notag\\
&&+\left. w_{3}\left( u_{2}u_{3}+u_{1}^{2}\right) \right]\notag \\
&&+\gamma _{UW,4}\left\{ u_{2}u_{3}\left( 2w_{1}-w_{3}\right)
+u_{1}u_{3}\left( 2w_{3}-w_{1}\right) \right.\notag \\
&&-\left. u_{1}u_{2}\left( w_{1}+w_{3}\right) \right\}\notag \\
&&+\kappa _{UW,4}\left\{ u_{3}\left( 2u_{3}w_{1}-u_{2}w_{3}\right)
+2u_{1}^{2}w_{3}\right.\notag \\
&&-\left. u_{1}\left( 2u_{3}w_{2}+u_{2}w_{1}\right) \right\}\notag \\
&&+4\gamma _{UW,5}w_{2}u_{2}^{2}+2\gamma _{UW,6}u_{2}\left(
u_{1}w_{1}+u_{3}w_{3}\right)\notag \\
&&+2\kappa _{UW,5}w_{2}u_{2}^{2}+2\gamma _{UW,7}\left(
u_{3}^{2}w_{1}+u_{1}^{2}w_{3}\right)\notag \\
&&+\frac{\kappa _{UW,6}}{2}\left[ w_{2}\left( u_{1}^{2}+u_{3}^{2}\right)
+u_{2}\left( u_{1}w_{1}+u_{3}w_{3}\right) \right]\notag \\
&&+\kappa _{UW,7}\left[ w_{1}u_{3}\left( u_{1}+u_{2}\right)
+w_{3}u_{1}\left( u_{2}+u_{3}\right) \right]\notag \\
&&+\kappa _{UW,8}\left[ w_{2}\left( u_{1}^{2}+u_{3}^{2}\right) -u_{2}\left(
u_{1}w_{1}+u_{3}w_{3}\right) \right]\notag \\
&&+\kappa _{UW,9}\left[ w_{1}u_{3}\left( u_{2}-u_{1}\right)
+w_{3}u_{1}\left( u_{3}-u_{2}\right) \right]\notag \\
&&+\kappa _{UW,10}w_{2}\left( u_{1}^{2}-u_{3}^{2}\right) ,
\end{eqnarray}
\begin{eqnarray}
\frac{\partial V_{U,W}}{\partial w\text{$_{_{3}}$}} &=&2\gamma
_{UW,1}w_{3}\left( u\text{$_{1}^{2}+u_{2}^{2}+u_{3}^{2}$}\right)\notag \\
&&+2\kappa _{UW,1}u_{3}\left( u_{1}w_{1}+u_{2}w_{2}+u_{3}w_{3}\right)\notag \\
&&+2\gamma _{UW,2}w_{3}\left( 2u_{3}^{2}-u_{1}^{2}-u\text{$_{2}^{2}$}\right)\notag
\\
&&+2\kappa _{UW,2}u_{3}\left( 2u_{3}w_{3}-u_{1}w_{1}-u_{2}w_{2}\right)\notag \\
&&+2\gamma _{UW,3}\left( w_{1}+w_{2}\right) \left[ u_{2}u_{3}+u_{1}\left(
u_{2}+u_{3}\right) \right]\notag \\
&&+2\kappa _{UW,3}\left[ 2w_{3}u_{1}u_{2}+w_{1}\left(
u_{1}u_{3}+u_{2}^{2}\right) \right.\notag \\
&&+\left. w_{2}\left( u_{2}u_{3}+u_{1}^{2}\right) \right]\notag \\
&&+\gamma _{UW,4}\left\{ u_{1}u_{3}\left( 2w_{2}-w_{1}\right)
+u_{1}u_{2}\left( 2w_{1}-w_{2}\right) \right.\notag \\
&&-\left. u_{2}u_{3}\left( w_{1}+w_{2}\right) \right\}\notag \\
&&+\kappa _{UW,4}\left\{ u_{1}\left( 2u_{1}w_{2}-u_{3}w_{1}\right)
+2u_{2}^{2}w_{1}\right.\notag \\
&&-\left. u_{2}\left( 2u_{1}w_{3}+u_{3}w_{2}\right) \right\}\notag \\
&&+4\gamma _{UW,5}w_{3}u_{3}^{2}+2\gamma _{UW,6}u_{3}\left(
u_{1}w_{1}+u_{2}w_{2}\right)\notag \\
&&+2\kappa _{UW,5}w_{3}u_{3}^{2}+2\gamma _{UW,7}\left(
u_{2}^{2}w_{1}+u_{1}^{2}w_{2}\right)\notag \\
&&+\frac{\kappa _{UW,6}}{2}\left[ w_{3}\left( u_{1}^{2}+u_{2}^{2}\right)
+u_{3}\left( u_{1}w_{1}+u_{2}w_{2}\right) \right]\notag \\
&&+\kappa _{UW,7}\left[ w_{1}u_{2}\left( u_{1}+u_{3}\right)
+w_{2}u_{1}\left( u_{2}+u_{3}\right) \right]\notag \\
&&+\kappa _{UW,8}\left[ w_{3}\left( u_{1}^{2}+u_{2}^{2}\right) -u_{3}\left(
u_{1}w_{1}+u_{2}w_{2}\right) \right]\notag \\
&&+\kappa _{UW,9}\left[ w_{1}u_{2}\left( u_{1}-u_{3}\right)
+w_{2}u_{1}\left( u_{3}-u_{2}\right) \right]\notag \\
&&+\kappa _{UW,10}w_{3}\left( u_{2}^{2}-u_{1}^{2}\right) ,
\end{eqnarray}
Considering the VEV configuration:
\begin{equation}
u_{1}=u,\hspace{0.3cm}u_{2}=u_{3}=0,\hspace{0.3cm}w_{1}=w_{2}=0,\hspace{0.3cm}w_{3}=w.
\label{VEVpattern2triplets}
\end{equation}
From the expressions given above, we find that the scalar potential minimization equations take the form:
\begin{eqnarray}
\frac{\partial V}{\partial u\text{$_{_{1}}$}} &=&\frac{u}{2}\left[ -4\mu
_{U}^{2}+8\left( \kappa _{U,1}+\kappa _{U,2}+\kappa _{U,5}\right)
u^{2}+\kappa _{UW,6}w^{2}\right.\notag \\
&&+\left. \left(4\gamma _{UW,1}-4\gamma _{UW,2}+2\kappa
_{UW,8}-2\kappa _{UW,10}\right) w^{2}\right]\notag \\
&=&0, \\
\frac{\partial V}{\partial u\text{$_{2}$}} &=&uw^{2}\left( 2\kappa
_{UW,3}-\kappa _{UW,4}\right) =0, \\
\frac{\partial V}{\partial u\text{$_{3}$}} &=&0, \\
\frac{\partial V}{\partial w\text{$_{1}$}} &=&0, \\
\frac{\partial V}{\partial w\text{$_{2}$}} &=&2u^{2}w\left( \gamma
_{UW,7}+\kappa _{UW,3}+\kappa _{UW,4}\right) =0, \\
\frac{\partial V}{\partial w\text{$_{3}$}} &=&\frac{w}{2}\left[ -4\mu
_{W}^{2}+8\left( \kappa _{W,1}+\kappa _{W,2}+\kappa _{W,5}\right)
w^{2}+\kappa _{UW,6}u^{2}\right.\notag \\
&&+\left. \left( 4\gamma _{UW,1}-4\gamma _{UW,2}+2\kappa
_{UW,8}-2\kappa _{UW,10}\right) u^{2}\right]\notag \\
&=&0.
\end{eqnarray}
Then, from the scalar potential minimization equations, we find the following relations:
\begin{eqnarray}
\kappa _{UW,4} &=&2\kappa _{UW,3}, \\
\gamma _{UW,7} &=&-\left( \kappa _{UW,3}+\kappa _{UW,4}\right) , \\
\mu _{U}^{2} &=&2\left( \kappa _{U,1}+\kappa _{U,2}+\kappa _{U,5}\right)
u^{2}+\left( 4\gamma _{UW,1}-4\gamma _{UW,2}\right.\notag \\
&&+\left. \kappa _{UW,6}+2\kappa _{UW,8}-2\kappa _{UW,10}\right) \frac{w^{2}%
}{4}, \\
\mu _{W}^{2} &=&2\left( \kappa _{W,1}+\kappa _{W,2}+\kappa _{W,5}\right)
w^{2}+\left( 4\gamma _{UW,1}-4\gamma _{UW,2}\right.\notag \\
&&+\left. \kappa _{UW,6}+2\kappa _{UW,8}-2\kappa _{UW,10}\right) \frac{u^{2}%
}{4}.
\end{eqnarray}
These results show that the VEV directions for the two $\Delta(27)$ triplets, i.e., $U$ and $W$ scalars in Eq. (\ref{VEVpattern2triplets}), are consistent with a global minimum of the scalar potential given in Eq. (\ref{scalarpotentialtwotriplets}) for a large region of parameter space. Furthermore, let us note that if one only considers one $\Delta(27)$ scalar triplet, by setting Eqs. (\ref{dV1})-(\ref{dV3}) to zero, it follows that the VEV pattern for the $\Delta(27)$ triplet $S$, pointing in the $(1,1,1)$ $\Delta(27)$ direction, is a natural solution of the scalar potential minimization equations.

\end{document}